\documentclass[nolinenumbers]{aastex631}

\usepackage{graphicx}	
\usepackage{amsmath}	
\usepackage{tablefootnote} 

\begin{document}

\title{Exploring the evolution of a dwarf spheroidal galaxy with SPH simulations: I. Stellar feedback}


\author{Roberto Hazenfratz}
\affiliation{Núcleo de Astrofísica, Universidade Cidade de São Paulo \\
R. Galvão Bueno 868, Liberdade, 01506-000\\
São Paulo, Brazil}

\author{Paramita Barai}
\affiliation{Istituto Nazionale di Astrofisica (INAF) \\ Osservatorio Astronomico di Trieste (OATs) \\
Trieste, Italy}

\author{Gustavo A. Lanfranchi}
\affiliation{Núcleo de Astrofísica, Universidade Cidade de São Paulo \\
R. Galvão Bueno 868, Liberdade, 01506-000\\
São Paulo, Brazil}

\author{Anderson Caproni}
\affiliation{Núcleo de Astrofísica, Universidade Cidade de São Paulo \\
R. Galvão Bueno 868, Liberdade, 01506-000\\
São Paulo, Brazil}


\begin{abstract}
A fundamental question regarding the evolution of dwarf spheroidal galaxies is the identification of the key physical mechanisms responsible for gas depletion. Here, we focus on the study of stellar feedback in isolated dwarf spheroidal galaxies, by performing numerical simulations using a modified version of the SPH code GADGET-3. The Milky Way satellite Leo II (PGC 34176) in the Local Group was considered as our default model dwarf galaxy. The parameter space for the stellar feedback models was explored to match observational constraints of Leo II, such as residual gas mass, total mass within the tidal radius, star formation history, final stellar mass, stellar ages and metallicity. Additionally, we examined the impact of the binary fraction of stars, initial mass function, dark matter halo mass and initial gas reservoir. Many simulations revealed recent star formation quenching due to stellar feedback. In general, the gas depletion, expected star formation history, total mass of stars and total mass within the tidal radius were adequately reproduced in the simulations when compared to observational estimates. However, there were discrepancies in the distribution of stellar ages and metallicities, which suggested that the cosmic gas infall would play a more complex role in our dwarf spheroidal galaxy than captured by a monolithic infall scenario. Our results suggest that currently quenched dwarf galaxies may not necessarily need to evolve within clusters or groups, and that stellar feedback alone could be a sufficient factor in shaping at least some of these galaxies as we observe them today. 

\end{abstract}


\section{Introduction}

In the hierarchical framework of structure formation in the universe, it is generally accepted that dark matter halos form in a sequence, with the least massive halos developing first and subsequently growing into more massive structures over cosmic time. This work focuses on the investigation of a dwarf spheroidal galaxy (dSph), which occupies the faint end of the galatic luminosity function \citep[e.g.][]{Mateo1998, Strigari2008, Grcevich2009, Mcconnachie2012}. These galaxies are typically spherical systems that are often devoid of gas, i.e. galaxies which are quenched at $z = 0$. Studying dwarf galaxies is of great interest because they make up the majority of galaxies in the universe and are abundant around the Milky Way (MW) and Andromeda (M31), thus influencing the evolution of more massive systems. Furthermore, they would be remnant galaxies of the first smaller blocks that gravitationally merged at high redshifts, eventually forming the large galaxies observed in the Local Universe \citep{Navarro1995, Moore1999, Robertson2005}. 

A fundamental question in understanding the evolutionary processes of dSph galaxies is the definition of the key mechanisms responsible for the exhaustion of their gaseous content. It encompasses the challenge of establishing the relative significance of environmental effects (such as ram pressure, tidal forces and reionization) versus the impact of internal feedback processes (such as supernova explosions) to shape the current stage of such systems. Although it is acknowledged that both environmental conditions and internal processes play pivotal roles in influencing many properties of dwarf galaxies \citep[e.g.][]{Higgs2021}, it is still unclear whether stellar feedback alone could shape field dwarfs or the most distant satellites the may have evolved independently from their host galaxies at earlier cosmic epochs.

Simulations of dwarf galaxies in the early Universe can be used to understand fundamental physical processes that govern galactic formation. The Local Group of galaxies is an excellent laboratory for providing observational constraints to these simulations, due to the great morphological variety and proximity of its satellite galaxies. This proximity facilitates the observations to obtain kinematic, dynamic and spectral information concerning the stars and interstellar medium in these galaxies (when detectable gas is present). In the past two decades, there has been a notable increase in research focused on simulating field and satellite dwarfs using both isolated \citep[e.g.][]{Valcke2008, Schroyen2011, Caproni2015, Vandenbroucke2016, Caproni2017, Emerick2019, Gutcke2021, Hislop2022} and cosmological \citep[e.g.][]{Kawata2006,Sawala2010, Revaz2012, Wheeler2015, Sawala2016, Wetzel2016, Revaz2018, Buck2019, Garrison2019, Wheeler2019, Fattahi2020, Rey2020, Jeon2021, Sanati2023} simulations. These studies have been unveiling the intricate and complex evolution histories of theses systems.

The galaxy used as a template in this work is the dSph Leo II (PGC 34176), which was discovered in 1950 by \citet{Harrington1950}. Its Galactocentric distance is estimated as $235.6^{+13.9}_{-9.14}$ kpc \citep{Li2021}, making it one of the most distant MW companions and therefore suitable for the study of internal feedback processes with relatively reduced influence from environmental effects, particularly in more recent stages of galactic evolution. \citet{Spencer2017} argue that Leo II can be considered a satellite galaxy of the Milky Way, based on its radial velocity and overall morphology. However, by analyzing the low value of the galactocentric radial velocity component and the absence of clear evidence for tidal disruption \citep{Koch2007, Coleman2007, Lepine2011}, it is possible to consider that this galaxy may have evolved in isolation and could be approaching the Milky Way for the first time in its evolutionary history. Additionally, \citet{Munoz2018} found that Leo II was the roundest dwarf galaxy in their MegaCam survey ($\epsilon = 0.07 \pm 0.02$), displaying a regular morphology without identifiable signs of tidal features. 
In a more recent study, \citet{Battaglia2022} conducted orbit determination of Local Group dwarf galaxies using \emph{Gaia} proper motions, estimating that Leo II passed at its pericenter ($\sim$ $40-200$ kpc) around 2 Gyr ago. Whether the satellite is in its first infall or heading out after its first pericentric passage, it would not significantly affect its star formation history, which was probably quenched before infall \citep{Dolphin2002, Koch2007, komiyama2007,Lanfranchi&Matteucci2010,Kirby2011multi}. The mean stellar age for Leo II was estimated as 9.4 Gyr by \citet{Dolphin2002} and as 8.8 Gyr by \citet{Orban2008}. 

Our primary objective was to assess whether the inclusion of winds driven by stellar evolution and SN in numerical simulations would enhance our ability to replicate key observational features and the star formation history of a dwarf spheroidal galaxy. To achieve this, we applied isolated galactic models to investigate the impact of star formation and its associated feedback mechanisms in the evolution of a system resembling the dSph Leo II regarding mainly the halo mass, star formation history, stellar mass and metallicity, morphology and relative isolation from massive hosts. Furthermore, we aim to verify whether stellar feedback alone could suffice in reproducing selected observational properties of the dSph Leo II and, consequently, of similar systems. It is essential to clarify that our primary objective is not to replicate Leo II with exactitude. Instead, this study aims to provide observational-motivated inputs and constraints for the application of phenomenological physical models in the examination of its galactic evolution.

Despite the clear advantages of using cosmological simulations to study galactic evolution within the cosmic framework of hierarchical formation of structures, isolated simulations offer the opportunity to rule out the effects of interest while minimizing the impact spurious perturbations. Isolated simulations also offer the flexibility to enhance the mass and spatial resolution, save computational time and facilitate a more straightforward analysis and interpretation of results \citep[e.g.][]{Pasetto2010}. Furthermore, these isolated simulations can yield valuable insights into optimizing parameters for cosmological zoom-in simulations by exploring the parameter space of subgrid models. 

As highlighted by \citet{Lanfranchi&Matteucci2010}, the study of a galaxy like Leo II can significantly contribute to our understanding of the formation and evolution of dwarf spheroidal galaxies that have evolved in relative isolation from larger hosts. Comparing the properties of these galaxies to those of nearby dwarf galaxies can help us disentangle the relative importance of stellar feedback and environmental effects in shaping their evolution. Understanding the mechanisms that lead to star formation quenching in dwarf galaxies is crucial for addressing the discrepancy between the observed number of galaxies orbiting the MW and Andromeda and the larger number of bound dark matter halos predicted by simulations of structure formation within the $\Lambda$CDM framework (Missing Satellites Problem) \citep[e.g.][]{Klypin1999, Bullock2017, Collins&Read2022}. 

Our study discusses the implications of both successful reproduction of observational constraints and remaining tensions, providing insights into the evolution of an isolated dwarf spheroidal galaxy. We deliberately maintained a simplified simulation framework to prioritize the investigation of internal effects of stellar feedback. Nevertheless, we recognize that the assembly of dwarf galaxies can be significantly more complex, involving factors such as multiple stellar populations, quenching by reionization and magnetic fields. Furthermore, this paper marks the beginning of a series of three. The second paper will explore AGN feedback from a putative intermediate-mass black hole within the same target galaxy, while the third paper will examine environmental effects through cosmological simulations and compare them with the findings obtained from the isolated scenario.

\section{Methods}

The hydrodynamical simulations to investigate stellar feedback in a dwarf spheroidal galaxy were performed with a modified version of the GADGET-3 code, which is based on smoothed particle hydrodynamics (SPH) for representation of fluids. These simulations are part of a larger project to explore different processes related to the formation and evolution of dwarf spheroidal galaxies in the Local Group. The project seeks to analyze the role and the interplay of internal feedback mechanisms and environmental effects for dwarf satellites of the Milky Way. This first paper, using the dwarf spheroidal Leo II as a reference, comprises the implementation of the stellar feedback and chemical evolution prescriptions of \citet{Tornatore2007} in an isolated dwarf galaxy.

The GADGET code was developed by \citet{Springel2001} with a Lagrangian approach suitable for both cosmological simulations, as well as isolated and merging/interacting structures \citep{Springel2001, Springel2005}. The code combines elements of SPH to model dynamical properties of gas, and N-body methods for non-collisional self-gravitating objects or fluids (dark matter, stars and black holes) to calculate the gravitational fields. In the GADGET-3 code an explicit entropy-conserving formulation is employed \citep{Springel2002}.

\subsection{Initial Conditions and Parameterization}

In our model galaxy, the dark matter distribution follows a Hernquist profile \citep{Hernquist1990}:

\begin{equation}
    \rho_{dm}(r)=\frac{M_{dm}}{2\pi}\frac{a}{r(r+a)^3}
\end{equation}

where $M_{\textit{dm}}$ is the total dark matter halo mass and $a$ is the scaling length of the profile. One of the motivations for using this profile is to represent an analytical distribution function that satisfactorily approximates the equivalent potential of Navarro-Frenk-White (NFW), presenting a steeper decline and mass convergence at large radii, which allows the construction of isolated haloes without the need for an ad hoc truncation.

The relationship between the two profiles is given by

\begin{equation}
    a=r_s\sqrt{2[\ln{(1+c)}-c/(1+c)]}
\end{equation}

where $c$ is a concentration index, given by $r_{200}/r_s$, with $r_s$ as the NFW profile scale length. 

The gas component is assumed as spherical for simplicity and is also modeled with a Hernquist profile:

\begin{equation}
    \label{eq:gas_ro}
    \rho_b(r)=\frac{M_b}{2\pi}\frac{b}{r(r+b)^3}
\end{equation}

where $b$ is the scale length of the gas profile (a free parameter). The gas mass is calculated as a fraction of the total mass: $M_b = m_b \times M_{\textit{tot}}$, where $M_{tot}$ is the total mass of the galaxy. 

The angular momentum of the DM profile is given by

\begin{equation}
    J=\lambda G^{1/2}M_{200}^{3/2}r_{200}^{1/2}\left(\frac{2}{f_c}\right)^{1/2}
\end{equation}

where $\lambda$ is a spin parameter and $f_c$ a factor that depends only on the concentration index $c$. The initial angular momentum is applied only to the dark matter particles. The value for $\lambda$ was chosen based on an average value found in literature \citep[][]{Bryan2013, Kurapati2018}.

The model and code used to generate initial conditions for isolated disc galaxies as described in \citet{Springel2005} was modified to represent dwarf spheroidal systems in the low-mass regime. The primary modifications included specifying the distinct concentration of the dark matter halos, the initial distribution of SPH gas particles, the initial gas velocity and the omission of the gas and stellar disc from the initial galaxy model. For the DM component initialization, the particle coordinates and velocities were drawn randomly from their respective distributions, the latter being approximated by a triaxial Gaussian. As for the spherical gas distribution, we adopted null initial velocities and $b = 50$ (in units of a virtual disk scale length, equivalent to $\sim$ 18.6 kpc) as a suitable scale length to enable smooth gas infall towards the center of the galaxy.

The free parameters of the model used for producing the initial conditions in the simulations of isolated dwarf galaxies were:\\

$v_{200}$: the circular velocity of the halo at the virial radius \citep{Strigari2007}.

$c$: concentration index \textbf{\color{blue} \citep{Correa2015, Cimatti2019}}.

$\lambda$: spin parameter \textbf{\color{blue} \citep[][]{Bryan2013, Kurapati2018}}.

$b$: scale length of the spheroidal distribution of gas.

$m_{b}$: mass fraction of the gas. 

The total initial mass of the galaxy can then be calculated by

\begin{equation}
    M_{tot}=\frac{v_{200}^3}{10GH_0}
\end{equation}

To properly account for gravitational softening lengths within the mass range of dwarf spheroidal galaxies, we employed the adaptive gravitational softening length method developed by \citet{Iannuzzi2011}. A reference value for the gravitational softening length, assuming $1/50$ of mean particle separation, would be 70 pc, which can be enhanced in the adopted scheme.

In addition to gravitational and hydrodynamic effects, radiative cooling processes are implemented for helium and hydrogen, combined with lines for radiative cooling referring to metals C, Ca, O, N, Ne, Mg, S, Si and Fe \citep{Tornatore2007, Wiersma2009}. The cooling tables represent a gas exposed to a redshift-dependent UV/X-ray background radiation from quasars and galaxies, based on the model of \citet{Haardt2001}, alongside the redshift-dependent cosmic microwave background radiation. The gas is assumed to be optically thin and in ionization equilibrium. 

The model for the isolated dwarf galaxies was first tested with gravity+hydrodynamics only (without cooling, star formation and feedback) to check for gravitational stability, The system was stable for approximately 13 Gyr.

\subsection{Multiphase model of star formation and thermal feedback}

Resolving spatial scales of molecular clouds targeting individual star formation remains computationally expensive, compounded by the incomplete details of star formation theory in the literature. In this study, we employ star formation and feedback recipes tailored to resolved scales ($\sim 10^4$ M$_{\odot}$ for gas). The star formation follows the effective subgrid scheme proposed by \citet{Springel2003} (abbreviated here as SH03), wherein the interstellar medium (ISM) is represented by a fluid formed by cold clouds ($T_c\sim1000$ K), confined and in pressure equilibrium with a hot ambient gas ($T_h\sim10^5-10^7$ K). The hydrodynamic equations are followed only by the ambient gas, while the cold clouds provide material for star formation and are subject to gravity, momentum transfer and participate in material and energetic exchanges with the ambient gas phase. All the the processes cited are computed for each particle in terms of simple differential equations that represent specific models for ISM physics \citep{McKee1977}. Motivated by the critical conditions for star formation observed by \citet{Kennicutt1989}, the occurrence of thermal instabilities is conditioned to a critical value of density, occurring for $\rho > \rho_{\textit{th}}$.

The representation of the star formation model and its thermal feedback in the SH03 approach consists of two main ingredients:

(i) A “law” or prescription for star formation, defining a rate motivated by the observations of \citet{Kennicutt1989, Kennicutt1998global}
\begin{equation}
    \frac{d\rho_*}{dt} = (1-\beta)\frac{\rho_c}{t_*}
\end{equation}

where $\rho_c$ is the density of cold clouds; $\beta$ is the fraction of short-lived stars that instantly die as supernovae; and $t_*$ is a characteristic time scale. For $t_*$, it is observed that an appropriate estimate, according to observations, is a proportional relation with the local dynamic time
\begin{equation}
    t_*(\rho) = t_*^0\left(\frac{\rho}{\rho_{th}}\right)^{-1/2}
\end{equation}

where $t_*^0$ is a parameter of the model, with the same fiducial value $t_*^0 = 1.5$ Gyr as assumed in \citet{Tornatore2007}; and $\rho$ is the total gas density, comprising the cold and hot phases ($\rho = \rho_c + \rho_h$).

(ii) An effective equation of state (EOS) for the ISM, given by
\begin{equation}
    P_{\textit{eff}} = (\gamma - 1)(\rho_hu_h + \rho_cu_c)
    \label{eq:eos}
\end{equation}

with $u$ referring to the specific thermal energy of the hot (\textit{h}) and cold (\textit{c}) phases; and $\gamma=5/3$ as the adiabatic expansion coefficient of the ideal monoatomic gas.

During the simulations, each star particle is treated as a simple stellar population, whose mass varies according the mass-dependent stellar lifetimes for a chosen initial mass function (IMF), accounting for mass losses. In this work, it was chosen the stellar IMF from \citet{Chabrier2003} for most of the simulations, within the mass range [0.1, 100] M$_\odot$. 

The SH03 model considers three basic processes responsible for mass exchange between phases: star formation, cold cloud evaporation by supernova feedback and cold cloud growth by radiative cooling. The adopted approach allows for the self-regulated treatment of star formation, by including stellar winds with the potential to cause its suppression over time. The energy balance for the gas can be written as

\begin{equation}
    \label{eq:energy_bal}
    \frac{d}{dt}(\rho_hu_h+\rho_cu_c)=-\Lambda_{net}(\rho_h,u_h)+\beta\frac{\rho_c}{t_*}u_{SN}-(1-\beta)\frac{\rho_c}{t_*}u_c
\end{equation}

In the balance of equation~\ref{eq:energy_bal}, $\Lambda_{net}$ represents the radiative cooling function of the hot ambient gas, which is the only component of the multiphase particle susceptible to radiative processes. The second term in the balance describes the injection of non-gravitational energy by the type II supernova explosion. Note that $\beta$ depends on the adopted stellar initial mass function (IMF), with $\beta \sim0.1$ for a Salpeter's IMF \citep{Salpeter1955}. The $u_{\textit{SN}}$ parameter is set to

\begin{equation}
    u_{SN} \equiv \frac{(1-\beta)}{\beta}\epsilon_{SN}
\end{equation}

whose value depends on the adopted IMF (see \autoref{sec:KinStellarFeed} for further details). Finally, the third term of the energy balance describes the energy that is lost from the gas phase by the material that is transformed into stars, whose conversion is assumed to occur at the temperature of cold clouds.

It is assumed that the energy resulting from the feedback of supernovae directly heats the hot phase, in addition to evaporating the cold clouds inside the hot bubbles of exploding supernovae by thermal conduction, returning material to the hot gas environment. 

\subsection{Kinetic stellar feedback model}\label{sec:KinStellarFeed}

To surpass the containment of the high entropy related to supernova remnants due to the constant coupling of hot and cold phases in the model SH03, an explicit wind model for stellar feedback is also considered in our simulations \citep{Springel2003, koudmani2022}.

The kinetic stellar feedback in the form of winds of supernovae is implemented using the energy-driven prescription and chemical evolution model from \citet{Tornatore2007}, which considers explicitly stellar lifetimes to determine the release of metals and energy, allowing the possibility to change the IMF and chemical yields as well. The wind mass-loss rate is expressed in function of the SFR as

\begin{equation}
    \label{eq:eta}
    \dot M_w = \eta \dot M_*
\end{equation}

where $\eta$ is the mass loading factor, which for disk galaxies may have a fiducial value $\eta = 2$ \cite[e.g.][]{Tornatore2007, Barai2013}, but which is higher in low-mass galaxies and needs to be further tested in the case of a dSph galaxy. 

The wind kinetic energy is considered to be a fixed fraction $\chi$ of the supernovae energy

\begin{equation}
    \label{eqn:chi}
    \frac{1}{2}\dot M_{w}v_{w}^2 = \chi \epsilon_{SN} \dot M_*
\end{equation}

where $v_{w}$ is the wind velocity and $\epsilon_{SN} = 1.1\times10^{49}$ erg M$^{-1}_{\odot}$ is the specific mean energy released by supernovae for the Chabrier IMF.

Equations \ref{eqn:chi} and \ref{eq:eta} can be easily combined to explicitly show the parameters to be tested ($\eta$, v$_{w}$) in a single equation

\begin{equation}
    \label{eqn:param_space}
    \frac{1}{2} \eta v_{w}^2 = \chi \epsilon_{SN}
\end{equation}

The parameter space to be tested corresponds to the interval $\eta = [5,1000]$ and $v_{w} = [20,350]$. To establish the lower limit tested for the wind mass loading factor, it was considered the value of $\eta = 2$ as the fiducial value used in cosmological simulations focusing on MW-type galaxies \citep[e.g.][]{Barai2013, Barai2018} and the minimum mass loading factors for the floor wind velocity imposed in the TNG simulations for halos with $M_h \lesssim 10^{11}$ M$_{\odot}$. On the other hand, some references for the upper limit of the mass loading factor were estimated by the best fits for $\eta_w \times M_{200c}$ for Illustris and TNG cosmological simulations \citep{Vogelsberger2014introducing, Vogelsberger2014properties, Pillepich2018}, considering a lower limit for the Leo II dark matter halo as $M_h \approx 4 \times 10^8$ M$_{\odot}$ \citep{Walker2007}.   

For the wind velocity, the lower limit estimation roughly corresponds to the minimum value of this velocity for the maximum value of the mass loading factor (eq.~\ref{eqn:param_space}), and the upper limit corresponds roughly to the wind velocity for the minimum value of mass loading factor, both for a fixed $\chi = 0.5$. Another basis for the choice of the minimum value can be found from the results of TNG, neglecting the wind velocity floor and extending the power law found ($v_{w} \propto M_h^{1/3}$). Furthermore, the value 350 km/s is the lower limit for wind velocities at injection in the TNG simulations, for dark matter halos with $M_h \lesssim 10^{11}$ M$_{\odot}$, adopted to enhance the stellar feedback efficiency to solve tensions between results and observations for halos less massive than those found in MW-type galaxies in the cosmological simulations \citep{Pillepich2018}. 

Currently, the model for star formation and feedback comprises the contribution of type Ia supernovae with $0.8 < M/$M$_{\odot} < 8$ \citep{Thielemann2003} and type II supernovae with $M/$M$_{\odot} > 8$ \citep{Woosley1995} for the energetic budget, stellar feedback and nucleosynthesis; and AGB stars \citep{VanDenHoek1997} solely for the nucleosynthesis. The nine chemical elements produced are C, Ca, O, N, Ne, Mg, S, Si and Fe. The mass-dependent time delays for different stellar populations to release metals was extracted from \citet{Padovani1993}. 

A fraction of the mass of a star particle is restored as diffuse gas during its evolutionary path and redistributed to the surrounding gas. The metal-enriched material and ejected energy are also spread among the neighbouring gas particles with weights determined by the SPH kernel. 

\subsection{Overview of the simulations}

Table~\ref{tab:config_table} shows the configuration and parameterization employed for simulating a dwarf spheroidal-like galaxy with the GADGET-3 code for $\sim$ 13.7 Gyr. The value for the virial velocity was extracted from the lower limit of the circular velocity estimated by \citet{Strigari2007}. Note the the value used differs from the value of $\sim$ 17 km/s found by the authors, because the main objective was reproducing the mass estimated for the galaxy at $z=0$, which is practically the dark matter halo mass, since no gas is currently observed for Leo II (but must be present at higher redshifts).  

\begin{table}[h]
	\centering
	\caption{Configuration and parameters for most of the SPH simulations of an isolated dwarf spheroidal galaxy, considering Leo II as a reference.}
	\label{tab:config_table}
	\begin{tabular}{lc} 
		\hline
		Parameter & Configuration/Value\\
		\hline
		$v_{200}$$^1$  & 20.5 km s$^{-1}$\\
		DM halo & Hernquist potential\\
		DM mass & $1.6 \times 10^9$ M$_{\odot}$\\
		DM concentration ($c$)$^2$ & 9\\
		Gas fraction ($m_b$) & 0.16\\
		Initial gas reservoir & $3.2 \times 10^8$ M$_{\odot}$\\
		Gas scale length ($b$) & 18.6 kpc\\
		Initial gas particle velocity & 0\\
		DM particle number  & 30000\\
		DM mass resolution & $5.3 \times 10^4$ M$_{\odot}$\\
		Gas particle number & 20000\\
	    Initial gas particle mass & $1.6 \times 10^4$ M$_{\odot}$\\
		Spin paramater ($\lambda$)$^3$ & 0.03\\
		Interpolation parameter ($f_{\text{eos}}$) & 1\\
		Gravitational softening length & adaptive\\
		Initial mass function & Chabrier / Salpeter\\
		\hline
		\end{tabular}
        \begin{minipage}{7cm}
        $^1$ \citet{Strigari2007}. For some simulations, the value of $v_{200} = 13$ km s$^{-1}$ was used instead. In this case, the dark matter halo mass is $\sim$4 $\times$ 10$^8$ M$_{\odot}$.\\
        $^2$ \citet{Correa2015, Cimatti2019}.\\
        $^3$ \citet{Bryan2013, Kurapati2018}.\\
        \end{minipage}
\end{table}

The value of the dark matter halo concentration was estimated as an intermediate value in the range $c \approx [5,13]$, for a DM halo mass compatible with Leo II ($\sim10^9$ M$_{\odot}$) in the interval of $z = [4,0]$, when most of the stars were formed in the universe. The curves for estimating $c$ were computed with a model based on the extended Press-Schechter formalism in \citet{Correa2015} and adapted by \citet{Cimatti2019}. Given that our idealized non-cosmological simulations lack the capability to track the evolution of the concentration parameter ($c$) over time, we opted to select a mean compromised value that reflects an average over the specified redshift interval. Note that starting with a higher value of $c$ and the entire current halo mass since earlier epochs could potentially diminish the influence of stellar feedback. The value for $m_b$ was chosen as the mean baryon fraction in the universe in most of the simulations \citep{Planck2020}. 


To establish the optimal initial conditions for simulating  an isolated dwarf spheroidal galaxy resembling Leo II, 82 short simulations (with 8000 DM particles + 5000 gas particles) were conducted to analyze physically plausible conditions. The optimal parameters were then selected for the long-run simulations described in Table~\ref{tab:simul_table}. 

\begin{table}[ht]
	\centering
        \caption{Parameter space for the long-run SPH simulations.}
        \label{tab:simul_table}
        \begin{tabular}{ccccccc} 
		\hline
	    Simulation & Label & $\eta$ & $v_{\textit{}{\text{wind}}}$ (km s$^{-1}$) & $\chi$ & BF & IMF\\
		\hline
		1 & $\eta$1000v25 & 1000 & 25 & 0.57 & 0.1 & Chabrier\\
		2 & $\eta$1000v350 & 1000 & 350 & 111 & 0.1 & Chabrier\\
		3 & $\eta$1000vCalc & 1000 & 24 & 0.50 & 0.1 & Chabrier\\
		4 & $\eta$5v350 & 5 & 350 & 0.56 & 0.1 & Chabrier\\
		5 & $\eta$500vCalc & 500 & 33 & 0.50 & 0.1 & Chabrier\\
		6 & $\eta$5vCalc & 5 & 333 & 0.50 & 0.1 & Chabrier\\
        7 & $\eta$5v200 & 5 & 200 & 0.18 & 0.1 & Chabrier\\
        8 & $\eta$50vCalc & 50 & 105 & 0.50 & 0.1 & Chabrier\\
        9 & $\eta$70vCalc & 70 & 89 & 0.50 & 0.1 & Chabrier\\
        10 & $\eta$60vCalc & 60 & 96 & 0.50 & 0.1 & Chabrier\\
        11 & $\eta$55vCalc & 55 & 100 & 0.50 & 0.1 & Chabrier\\
        12 & $\eta$60vCalcBF4 & 60 & 96 & 0.50 & 0.4 & Chabrier\\
        13 & $\eta$60vCalcBF4-v2 & 60 & 96 & 0.50 & 0.4 & Chabrier\\
        14 & $\eta$60vCalcBF4-Sal & 60 & 96 & 0.50 & 0.4 & Salpeter\\
        15 & $\eta$60vCalcBF4-HR$^1$ & 60 & 96 & 0.50 & 0.4 & Chabrier\\
        16 & $\eta$30vCalcBF4-Sal & 30 & 136 & 0.50 & 0.4 & Salpeter\\
        17 & $\eta$30vCalcBF4-LM$^2$ & 30 & 136 & 0.50 & 0.4 & Chabrier\\
        18 & $\eta$60v117BF4 & 60 & 117 & 0.75 & 0.4 & Chabrier\\
		19 & $\eta$70v117BF4 & 70 & 117 & 0.87 & 0.4 & Chabrier\\
	    20 & $\eta$60vCalcBF4-LM & 60 & 96 & 0.50 & 0.4 & Chabrier\\
        21 & $\eta$45vCalcBF4-Sal & 45 & 111 & 0.50 & 0.4 & Salpeter\\
        22 & $\eta$45v135BF4-Sal & 45 & 135 & 0.75 & 0.4 & Salpeter\\
        23 & $\eta$60vCal-b5 & 60 & 96 & 0.50 & 0.1 & Chabrier\\
        24 & $\eta$45vCalc-b30 & 60 & 96 & 0.50 & 0.1 & Chabrier\\
        25 & thermal-only & - & - & - & 0.1 & Chabrier\\
        \hline
        \end{tabular}
        \begin{minipage}{7cm}
        $^1$HR - high resolution.\\
        $^2$LM - low mass (for the dark halo).
        \end{minipage}
\end{table}

\pagebreak
\section{Results and Discussion}

In this section, we first present an optimal simulation serving as a reference run for further analysis, along with the constraints employed in its selection. Following this, we investigate the impact of the parameter space of the kinetic stellar feedback model and specific aspects of the galaxy setup on its evolution. 

\subsection{Fiducial simulation} \label{optsim}
\label{subsec:leo}

\begin{figure*}
    \includegraphics[width=\textwidth]{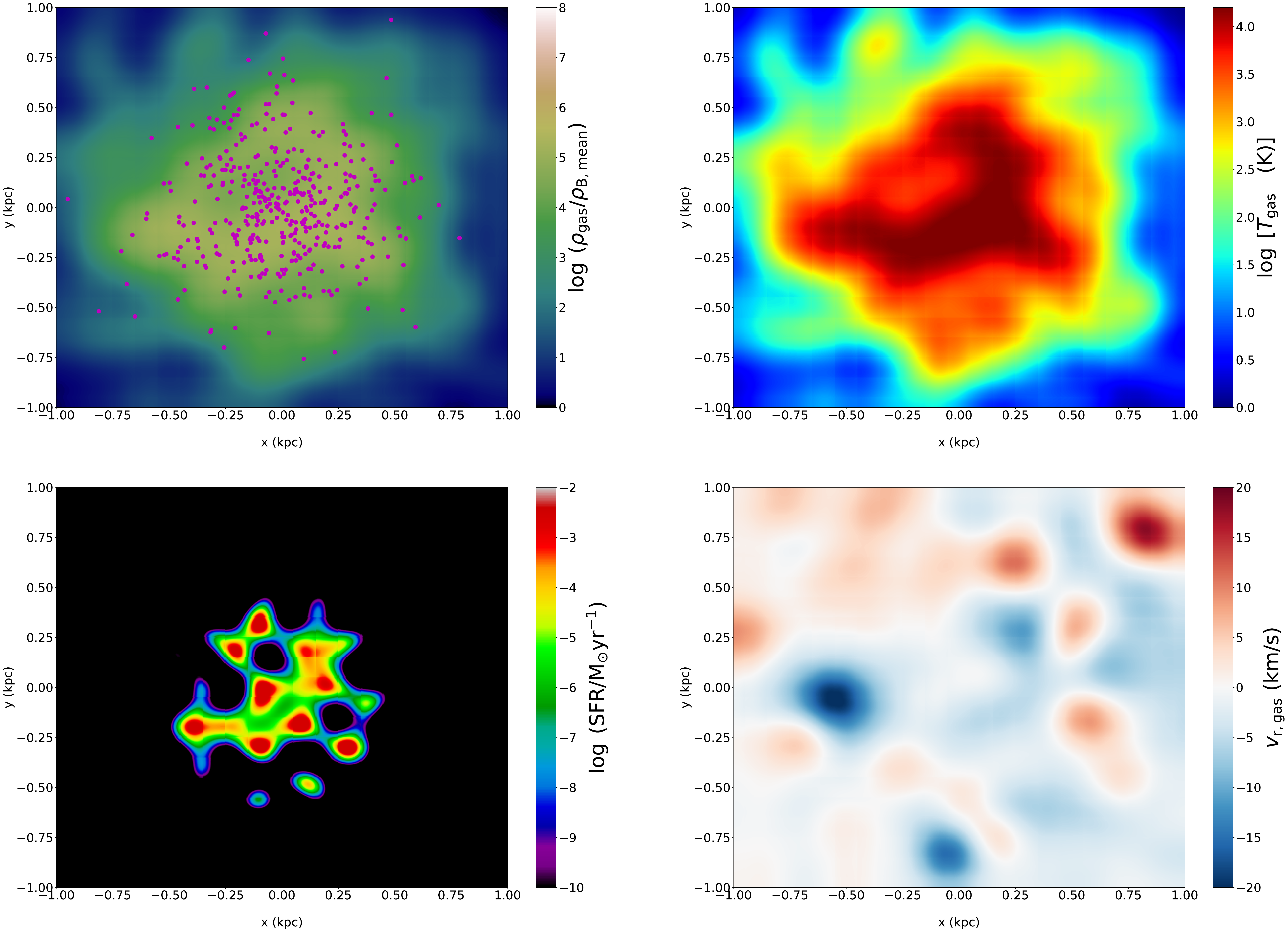}
    \vspace*{2mm}
    \caption{Maps in the xy-plane for overdensity (top left), temperature (top right), star formation rate (bottom left) and radial velocity (bottom right) at t = 1.96 Gyr for the fiducial simulation of Leo II ($\eta$60vCalc). The magenta dots on the overdensity map represents the position of star particles.}
    \label{fig:2d_maps}
\end{figure*}

We select here an fiducial isolated simulation that best aligns with chosen constraints for modeling a dwarf spheroidal galaxy with characteristics similar to Leo II. Our selection criteria were based on the total gas depleted within the observationally estimated tidal radius, the star formation duration, the total stellar mass and median stellar metallicity as the primary constraints.  In this regard, the simulation $\eta$60vCalc (simulation 10) was chosen to be further explored in the following sections, emphasizing its satisfactory fits and tensions. This simulation features a mass loading factor of $\eta = 60$, and wind kick velocities calculated by Eq.~\ref{eqn:param_space}, with $\chi = 0.5$ ($v_{\textit{wind}} \sim96$ km s$^{-1}$). Fig.~\ref{fig:2d_maps} provides an example of a 2D map illustrating the overdensity, temperature, star formation rate and radial gas velocity (along versor $\hat{r}$ in spherical coordinates) for a cut in the xy-plane for the selected fiducial simulation.

\subsubsection{Mass evolution}

\begin{figure*}
    \includegraphics[width=\textwidth]{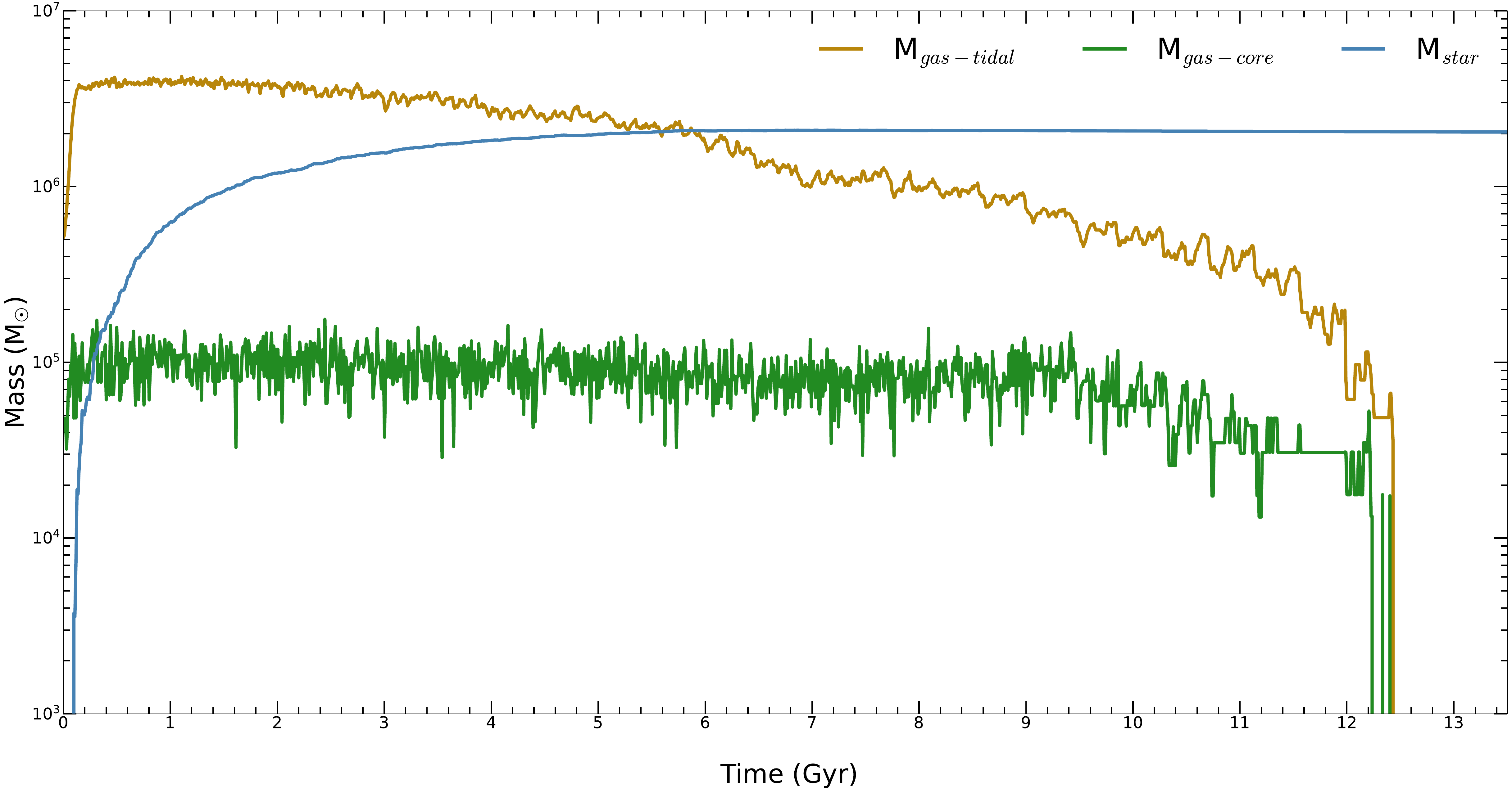}
    \vspace*{0mm}
    \caption{Mass evolution of different components for an isolated galaxy simulation using the dSph Leo II as reference. The parameters for the the kinetic stellar feedback model are $\eta = 60$ and $v_{\textit{wind}} = 96$ km s$^{-1}$.}
    \label{fig:mass}
\end{figure*}

The mass evolution for gas and stars is presented in Fig.~\ref{fig:mass}. The gas is divided into two components: 'core' for the gas within the current core radius of Leo II ($\sim180$ pc); and 'tidal' for the gas within the current tidal radius of the galaxy ($\sim650$ pc) \citep{Coleman2007}. 


Within the core region (180 pc), the general trend is an increase in the gas mass up to $\sim$10$^5$ M$_{\odot}$ until $\sim$100 Myr, followed by a decrease until 12.4 Gyr, although it is not strictly monotonic, as there are alternate periods of inflow and outflow of gas, with oscillations $\lesssim$ 0.8 dex. These patterns can be attributed to the continuous injection of energy from stellar feedback over time, which counterbalances the gravitational pull upon the gas.

For the tidal region (650 pc), the general trend is an increase in the gas mass until $\sim200$ Myr up to $\sim4 \times 10^6$ M$_{\odot}$, which is related to gas falling into the gravitational potential of the galaxy. From $\sim500$ Myr on, it starts a continuous depletion of gas until $\sim12.4$ Gyr, when the tidal region is finally completely depleted of gas. This intense gas depletion is an important feature to replicate in dwarf spheroidal galaxies. Specifically for Leo II, the estimated upper limit for detectable gas in the galaxy would be $\sim 10^4$ M$_{\odot}$ \citep{Grcevich2009}. 

Both in the tidal and core regions, the sudden drop of gas that happens after 12 Gyr is related to the moment when the mass of the residual gas is comparable to the mass of SPH particles in our simulations ($\sim10^4$ M$_{\odot}$). Once the last particle outflows, the gas mass drops to zero within the considered volumes. 

Particularly, the increase in gas mass until $\sim$ 200 Myr can be correlated with the rise in the star formation rate prior to its peak at $\sim10^{-3}$ M$_{\odot}$ yr$^{-1}$ around 600 Gyr in Fig.~\ref{fig:sfr}. During this process, the short-living stars inject thermal and kinetic energy in the gas, which is sufficient to generate a general trend of outflow for $t \gtrsim 200$ Myr in the core and tidal spheres, although still insufficient to completely remove the gas from these regions. 


Regarding the stellar mass formed during the galaxy evolution, it can be observed in Fig.~\ref{fig:mass} that most of the buildup of mass is concentrated in the first $\sim$4 Gyr, although the star formation proceeds at lower rates until $\sim$11.4 Gyr, when the last star formation episode happens (see Fig.~\ref{fig:sfr}). 

\citet{Kirby2011multi} argued that due to its distance and unknown orbit, Leo II could have spent most of its time in a low-density region of the Local Group, escaping disruptive gravitational interaction and gas stripping from the MW. But the lower density of such a region also poses a challenge for their interpretation of the metallicity distribution function, which suggests an increase in the gas reservoir. In the simulation of this work, although the total amount of gas within the tidal and core regions diminishes over time, it is not totally depleted until at least 12.4 Gyr (Fig.~\ref{fig:mass}). Therefore, this gas reservoir would provide the fuel for the sustained star formation observed in Fig.~\ref{fig:sfr}, without the need for additional inflows by interactions with other systems. However, the detailed dynamics of the inflowing gas might still be inaccurate to reproduce the metallicity patterns expected for Leo II, as discussed in sections \ref{chem_grad} and \ref{metallicity}.

It was calculated that the mass within a region of 600 pc (a common radius for comparing Local Group dwarfs) for the fiducial simulation is $3.7 \times 10^7$ M$_{\odot}$, a value consistent with the upper estimate of $2.1^{+1.6}_{-1.1} \times 10^7$ M$_{\odot}$ from \citet{Strigari2007}, or with $2.8 \times 10^7$ M$_{\odot}$ from \citet{Walker2007}. Note that this value depends essentially on the dark matter halo mass, halo concentration and stellar mass within the tidal region (with no gas detected).

\subsubsection{Star formation}

The graph depicting the star formation rate over time for the fiducial simulation is presented in Fig.~\ref{fig:sfr}. The star formation history (SFH) of the galaxy demonstrates the self-regulation achieved by the model for star formation and feedback. It can be roughly divided into three stages:

\begin{figure*}
    \includegraphics[width=\textwidth]{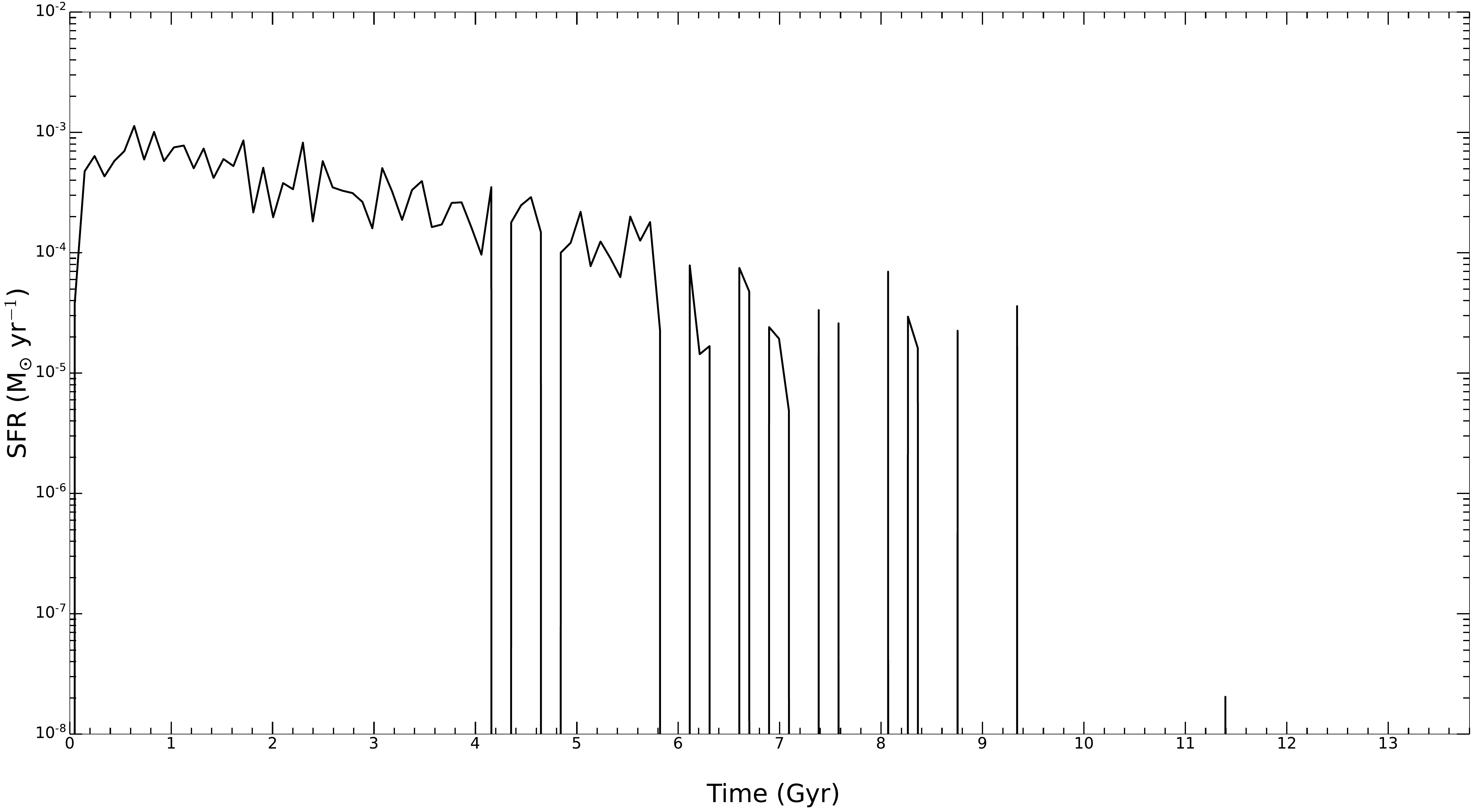}
    \vspace*{0mm}
    \caption{SFH for the fiducial simulation of Leo II ($\eta$60vCalc).}
    \label{fig:sfr}
\end{figure*}

(i) a continuous star formation stage until $\sim$4.2 Gyr. Its ascending phase lasts until $\sim$0.6 Gyr (with a peak at $\sim$10$^{-3}$ M$_{\odot}$ yr$^{-1}$) and its descending phase lasts from this point until $\sim$4.2 Gyr. For most of the period, the SFR is between $\sim$10$^{-4}$ and $\sim$10$^{-3}$ M$_{\odot}$ yr$^{-1}$.

(ii) a stage of more episodic star formation between $\sim$4.2 and $\sim$7.1 Gyr. The maximum time span between two consecutive active periods can be estimated as $\sim$300 Myr. The duration of most episodes is $0.1 \lesssim t_{\text{SF}} \gtrsim 1$ Gyr. 

(iii) a last phase with single bursts of star formation, with exception for an episode  between 8.2 and 8.4 Gyr, which lasts $\sim$100 Myr. The last single burst happens at $\sim$11.4 Gyr, after a period of 2 Gyr of no star formation.

These results are in general agreement with the star formation duration of 7 Gyr predicted by \citet{Lanfranchi&Matteucci2010} and \citet{Kirby2011multi}, which here would comprise phases (i) and (ii). The single bursts after 7 Gyr are also plausible, since there is evidence of intermediate-age stellar populations in the central region of the galaxy, with the stellar age tracers indicating an extended star formation history for Leo II, but with little SF activity in the last 6 Gyr \citep{Aaronson1983, Aaronson1985, Mighell1996, Koch2007, komiyama2007}. Furthermore, the duration and overall tendency of the SFR curve reasonably agrees with that obtained in \citet{Dolphin2002}, where the SFH were obtained by CMD numerical analysis for Leo II.

The final stellar structure of Leo II extends beyond the tidal radius of the system, roughly up to 3 kpc (maximum radial distance for a star particle $\sim5.3$ kpc at 13.7 Gyr), which aligns with evidence found by \citet{komiyama2007} of an extended stellar halo, formed by extra-tidal stars gravitationally bound to the galaxy. The authors even found a knotty substructure at the extended halo, which could be a small globular cluster that disrupted in the past and merged with the main body of Leo II. The resolution of stellar particles in this work ($\sim$10$^3$ M$_{\odot}$) and the isolated nature of the simulations limit any further comparison in this context. Future cosmological simulations may provide more insight into these issues.   

Concerning the more intermittent period of star formation after 7 Gyr, there is evidence in the literature that the MW dwarfs experienced alternating periods of increase and decrease in the SFR \citep[e.g.][]{Koch2007, Kirby2011metals}, which could be correlated to alternating periods of inflow and outflow of gas. 

In our simulations, the star formation quenching until recent times can be attributed to the continuous input of thermal and kinetic energy by stellar feedback, rather than a lack of gas for star formation. This is a plausible result in low-mass galaxies \citep[e.g.][]{Wetzel2015}. Furthermore, similar tendencies were also observed in other simulations (eg.: $\eta$1000v25, $\eta$1000vCalc, $\eta$500vCalc), where the star formation was sustained for 13.5 Gyr with specific combinations of wind mass loading factor and velocity. 

\begin{figure*}
    \includegraphics[width=\textwidth]{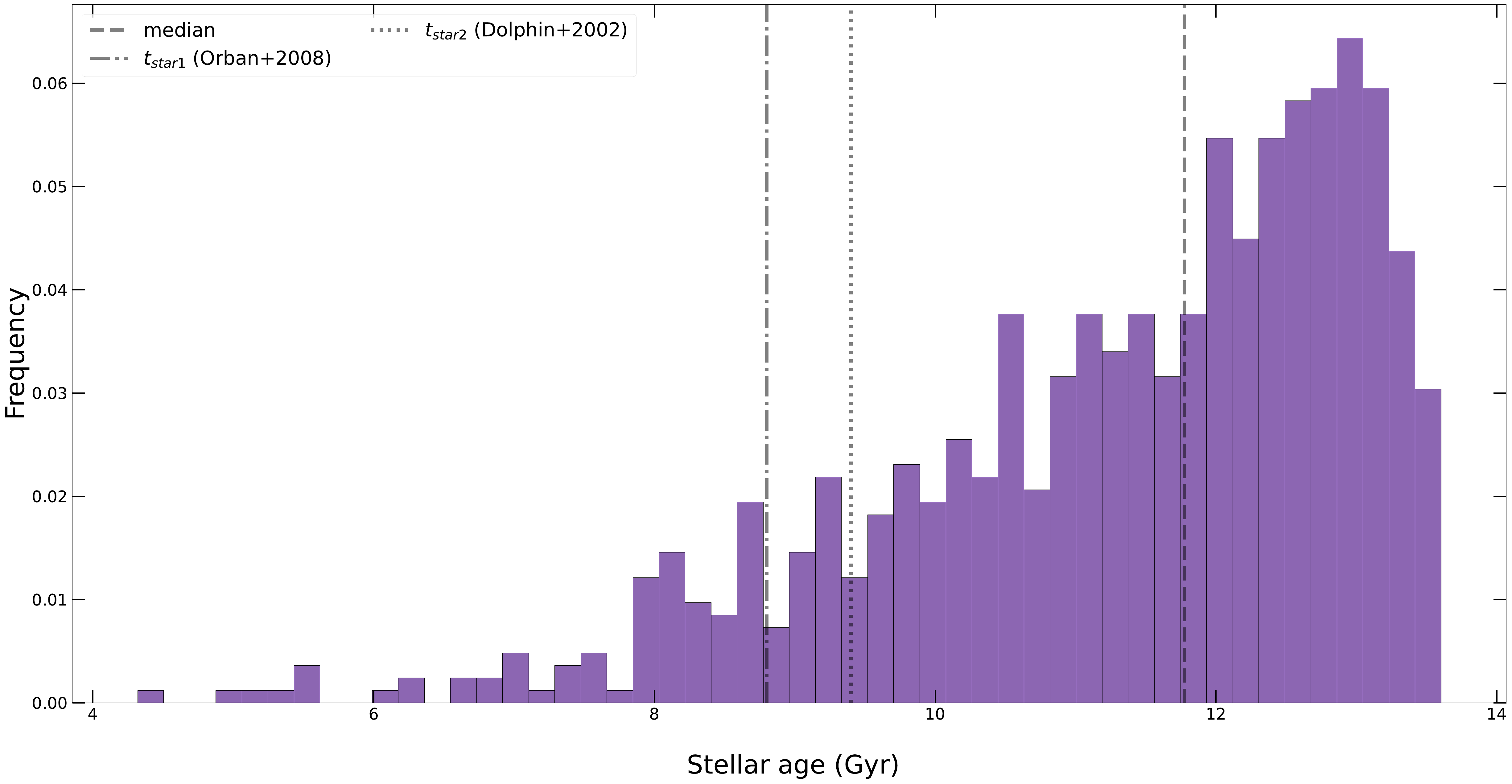}
    \vspace*{2mm}
    \caption{Distribution of stellar ages at t = 13.7 Gyr for the fiducial simulation of Leo II ($\eta60$vCalc).}
    \label{fig:star_ages}
\end{figure*}

The star formation history shown in Fig.~\ref{fig:sfr} generates the distribution of stellar ages presented in Fig.~\ref{fig:star_ages}, with a median age of $\sim$11.8 Gyr. This value is higher the one predicted by \citet{Orban2008} as 8.8 Gyr, or by \citet{Dolphin2002} as 9.4 Gyr, indicating the presence of older stars. It might reflect an inaccurate and faster gas infall over the evolution of the galaxy, which could generate chemically richer (as will be discussed in sections \ref{metallicity} and \ref{chem_grad}) and older stars than what is expected by observational constraints. Note that an intermittent regime of gas infall with more pristine composition could potentially address this tension, as already suggested by \citet{Kirby2011multi} for the dSph Leo II. 

It was observed that the star formation is primarily concentrated within the tidal region of Leo II. As an example showing the location of star formation, Fig.~\ref{fig:2d_maps} shows a cut in the xy-plane mapping the star formation rate for $t = 0.98$ Gyr. Furthermore, in progressive plots of the SFR over time, it was verified that the star formation becomes more centrally concentrated (although always inside 750 pc in the xy plane), which agrees with the conjecture for the evolution of Leo II in \citet{komiyama2007} regarding more recent star formation in the central regions of the galaxy.

\begin{figure*}
    \includegraphics[width=\textwidth]{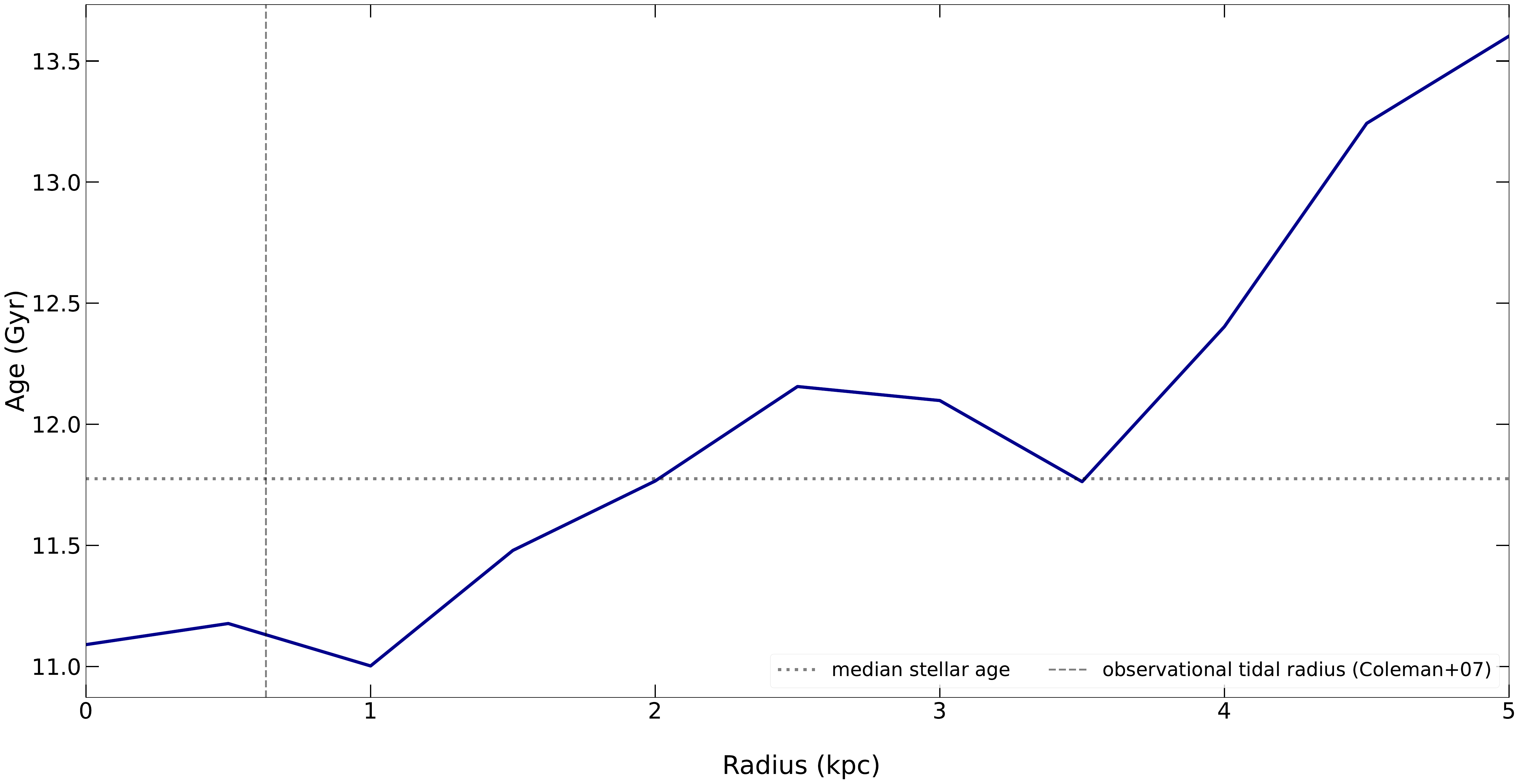}
    \vspace*{0mm}
    \caption{Stellar age profile at 13.7 Gyr for the fiducial simulation of Leo II dSph ($\eta$60vCalc).}
    \label{fig:stellar-age-profile}
\end{figure*}

To investigate the presence of any stellar age gradient, as also identified in \citet{komiyama2007}, a radial profile of this variable was plotted in Fig.~\ref{fig:stellar-age-profile}. The estimated stellar ages are lower in the central region when compared to larger distances. It can be associated to stellar migration over time, since all the SF in the simulations was observed within the central kpc (example in Fig.~\ref{fig:2d_maps}). Furthermore, the only two stellar particles with ages lower than 6 Gyr were found within 1 kpc: one between the core and tidal radius and the other outside the tidal radius.

A stronger stellar feedback could in principle generate more powerful outflows to delay the peak of star formation. However, simulations such as $\eta$60v117BF4, $\eta$60v117BF4-Sal and $\eta$45v135BF4-Sal, which involve a change in the fraction of the supernovae energy coupled to stellar winds (factor $\chi$ in Eq.~\ref{eqn:chi}, with values in Tab.~\ref{tab:simul_table}) and in the IMF used (Chabrier to Salpeter), still generated higher medians for the stellar ages. These more powerful winds suppressed the SF, producing lower stellar masses (Tab.~\ref{tab:metals}) and shorter periods of star formation. So, based on these results, the assumption of stronger stellar feedback could be ruled out.

Alternatively, environmental effects (ram pressure and/or tidal interactions) could generate turbulence in the ISM during early times and partially remove the gas supply prone to star formation, leading to a milder slope for the SFR. However, as discussed earlier, there is evidence against this hypothesis in the literature \citep{Koch2007, Coleman2007, Munoz2018}. On the other hand, the substructure identified by \citet{komiyama2007}, as potential debris of a globular cluster in Leo II, may suggest tidal interactions of this dwarf galaxy with smaller systems in the past (see section \ref{metallicity}).

\subsubsection{Radial profiles of gas properties} \label{gas-profiles}

\begin{figure*}
    \includegraphics[width=\textwidth]{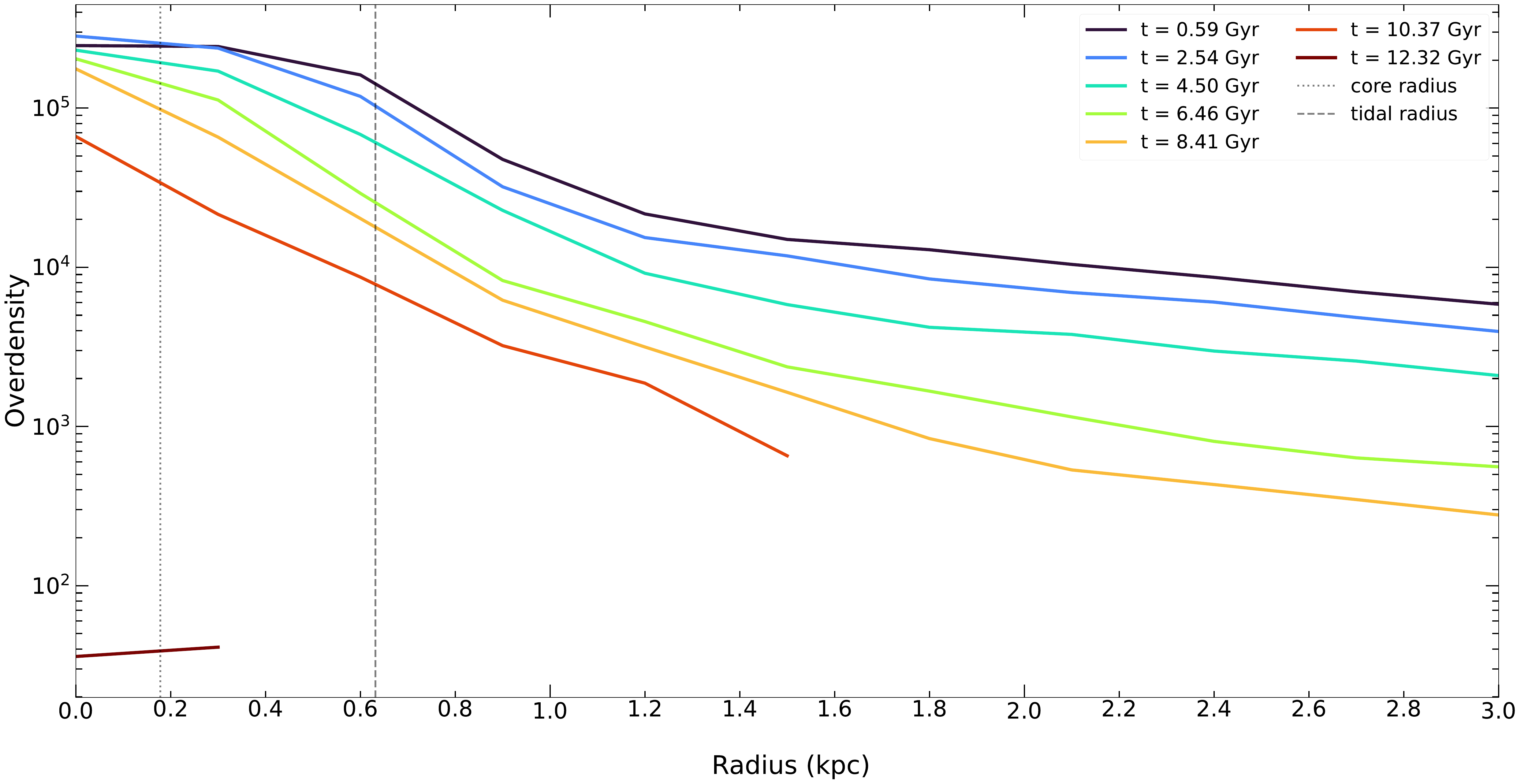}
    \vspace*{2mm}
    \caption{Radial distribution of overdensity for the fiducial simulation of Leo II ($\eta60$vCalc)}
    \label{fig:dens_prof}
\end{figure*}

\begin{figure*}
    \includegraphics[width=\textwidth]{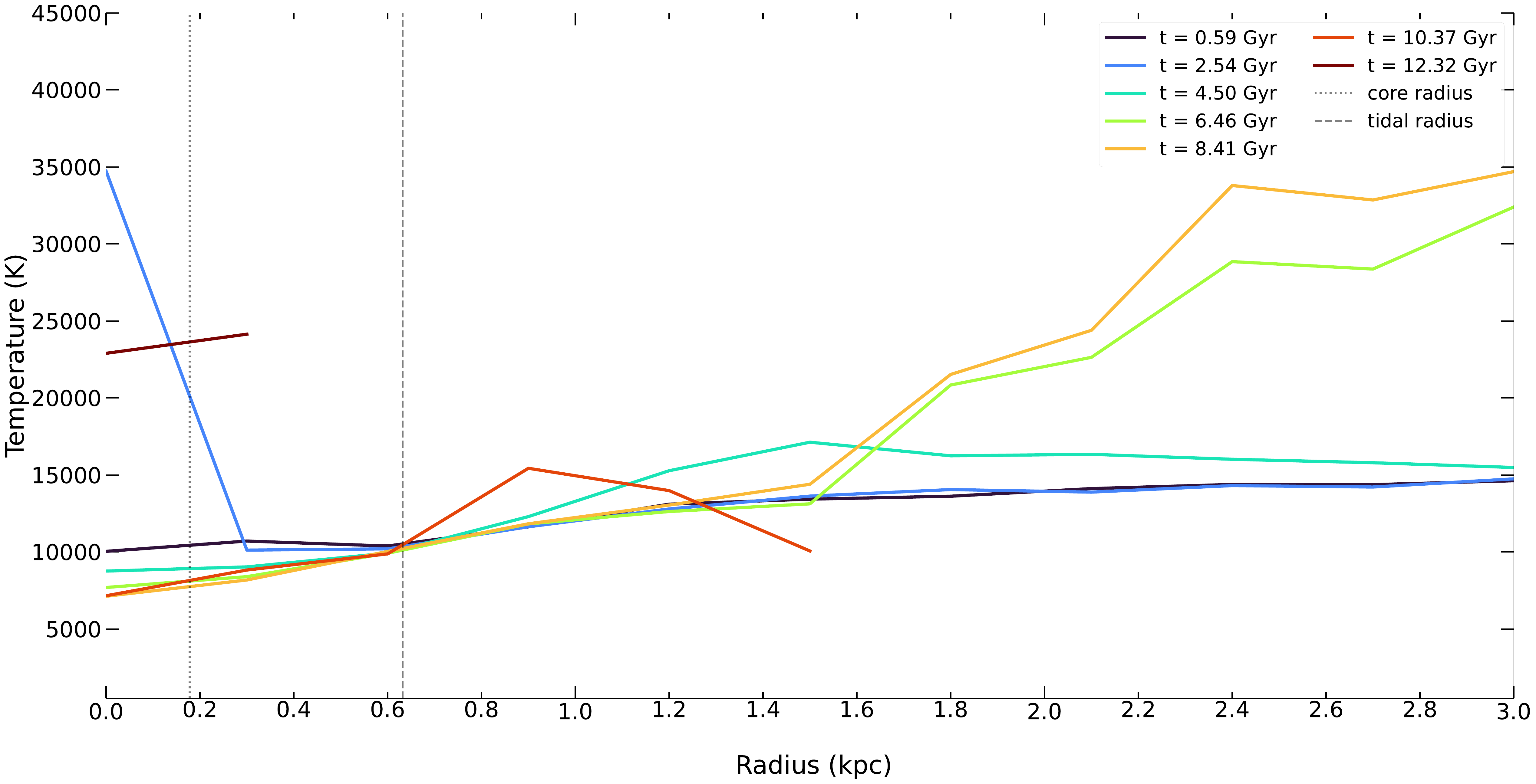}
    \vspace*{2mm}
    \caption{Radial distribution of temperature for the fiducial simulation of Leo II ($\eta60$vCalc)}
    \label{fig:temp_prof}
\end{figure*}

Figs.~\ref{fig:dens_prof} and ~\ref{fig:temp_prof} depict the variation of the gas overdensity and temperature with the galactic radius at selected times. 

In general, the overdensity decreases with radius and time in Fig.~\ref{fig:dens_prof}, indicating a general trend of gas depletion in the region within 3 kpc from the galactic center, which can be associated to the cumulative effect of energetic input by stellar feedback over the galaxy evolution. This trend was also observed in the 2D maps for overdensity in Fig.~\ref{fig:2d_maps} for different times. The steepness of the curves are higher within the central kpc compared to higher radii. It comprises the region where most of the stars are formed (see SF location example in Fig.~\ref{fig:2d_maps}), and reflects the general trend of gas depletion identified for the core and tidal regions in Fig.~\ref{fig:mass}. The only exception can be identified by comparing curves for 0.59 and 2.54 Gyr in the central 300 pc, where a higher overdensity is associated with a specific moment of gas infall in that region. This corresponds to alternating episodes of gas inflows and outflows, as identified in the mass curves within the tidal and core radius in Fig.~\ref{fig:mass}.

Regarding the temperature profiles in Fig.~\ref{fig:temp_prof}, the patterns are more complex than those for overdensity. Within the tidal region, the temperature rises more than 30000 K between $t = 0.59$ and $t = 2.54$ Gy, which can be associated to the SF that reaches its peak during this period (Fig.~\ref{fig:sfr}). This variation in the temperature can be explained by the thermal energetic input and by the thermalization of kinetic energy from gas particles forming the galactic winds. However, temperature tends to decrease afterwards, until $t = 10.37$ Gyr, as the SF tends to cease. The exception for curve at $t = 12.32$ can be associated to a few gas particles affected by the energetic input from SNe Ia (since SF has already ceased) or wind particles which were thermalized and pushed back to the central regions. Note that this curve does not continue at higher radii, as the gas is depleted over time, preventing further analysis.  

Outside the galactic central kiloparsec, the temperature increases with radius, reaching progressively higher temperatures at larger radii over time between $t \sim[4.5, 8.4]$ Gyr. This trend can be associated with a continuous input of feedback energy by stars at progressively higher radii and to wind particles flowing from more central regions to reach the galactic outskirts over time. During their trajectory outwards, these gas particles thermalize their kinetic energy by shock-heating and turning into hotter, low-density gas in the outer regions of the galaxy (see a similar discussion in \citet{Barai2018} for larger galaxies). Regarding an apparent inversion in this trend for 10.37 Gyr, note that this curve, like the last one, is not continuous, as progressively less gas particles are present in the central regions over time (due to the gas depletion observed in Fig.~\ref{fig:mass}), lacking sufficient statistics for further analysis. In fact, considering the same temperature profile for larger radii, it was be observed that the curve for this snapshot follows the same tendency identified for the others, reaching higher temperatures at higher radii.

\subsubsection{Outflows}

The 2D map of gas radial velocity for $\sim$ 1 Gyr in Fig.~\ref{fig:2d_maps} shows the presence of gas outflows (red) and inflows (blue), with a relative predominance of inflows in regions where the SF is taking place. The absolute values of the radial velocities are $\lesssim 20$ km.s$^{-1}$ for the central region depicted in the plot. 

The number fraction of the outflowing gas particles for the simulation $\eta$60vCalc was calculated as 0.67 at 12 Gyr (Table \ref{tab:metals}). This value indicates that most of the gas particles in the simulation at this time, when the star formation has already ceased (Fig.~\ref{fig:sfr}), can be classified as outflowing particles, which are related to the gas that can escape the galactic gravitational potential. This trend is related to the energy input by stellar feedback, which, if not present, could not prevent the gas from falling continuously into the gravitational potential of the galaxy. 

\begin{figure*}[ht]
    \includegraphics[width=\textwidth]{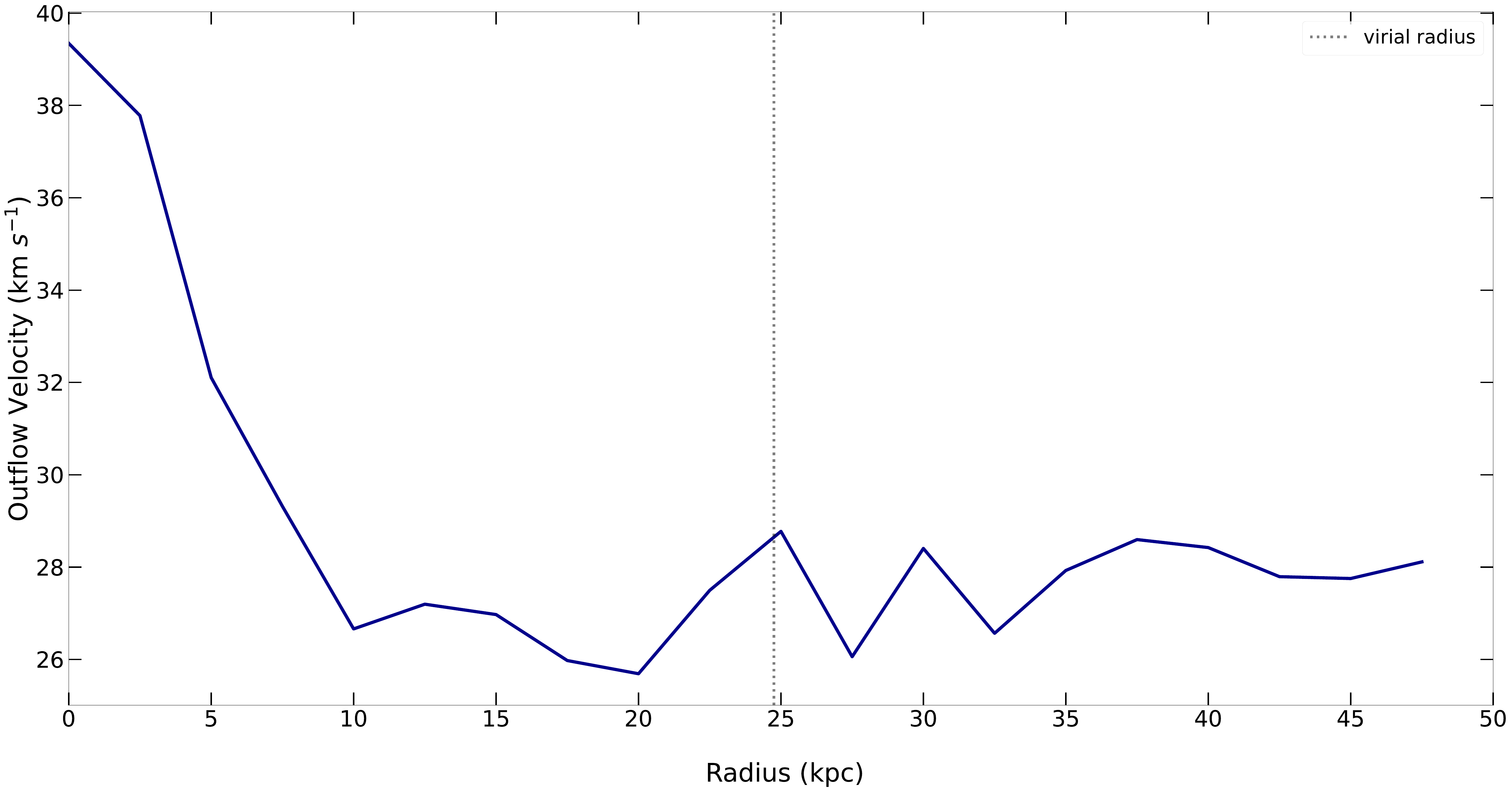}
    \vspace*{2mm}
    \caption{Radial distribution of gas outflow radial velocities for the fiducial simulation of Leo II ($\eta60$vCalc) at 4.01 Gyr.}
    \label{fig:outflow-vel}
\end{figure*}

Fig.~\ref{fig:outflow-vel} illustrates the radial profile of velocities for the outflowing gas particles in the fiducial SPH simulation at $\sim 4$ Gyr, which represents approximately the end of the first and continuous period of star formation in Fig.~\ref{fig:sfr}. The velocities fall within the interval of 25-40 km/s. This range aligns with the velocities typically observed for large-scale gaseous outflows in disky dwarf galaxies, which are the type of dwarf galaxy with more observational constraints due to the current observational challenges. In such galaxies, the galactic winds usually have a conical shape and reach velocities of 25-100 km/s \citep{Collins&Read2022}. One striking difference is that in disky dwarfs these winds move perpendicular to the stellar and gaseous disk, whereas in our simulations the outflows are isotropic by construction. 

The reduction in gas outflow velocity with galactic radius, up to $\sim$ 10 kpc, can be attributed to the increasing gas temperature over time and with distance from the center, as discussed in section~\ref{gas-profiles} with Figs.~\ref{fig:temp_prof}. The gas particles flowing outward experience a reduction in their velocities due to collisions with other gas particles, partially thermalizing their kinetic energy along their pathway out of the galaxy.

The outflowing gas mass rates for the fiducial simulation are plotted in Fig.~\ref{fig:Mout-thermal} for $t = 6.85$ Gyr. A gas particle was classified as an outflowing particle if its velocity was greater than the escape velocity of the galaxy ($v_r > v_{\textit{esc}} \sim24$ km s$^{-1}$). The outflow rates were then calculated as in \citet[][]{Barai2018}

\begin{equation}
    \label{eqn:outflow-rate}
    \dot M_{out}(r) = \sum_{v_{r,i} > v_{esc}} \frac{m_i |v_{r,i}|}{\Delta r} 
\end{equation}

where $m_i$ is the mass of a gas particle \textit{i}, $v_{r,i}$ is the radial component of the gas particle velocity, and $\Delta r$ is the width of the spherical shell used in the calculation. In Fig.~\ref{fig:Mout-thermal}, the local outflow rates are located in the range $2 \lesssim \dot M_{out} \lesssim 9$ M$_{\odot}$ yr$^{-1}$ along the galactic radius. 

\subsubsection{Chemical abundance distributions and gradients} \label{chem_grad}

\begin{figure*}[ht]
    \includegraphics[width=\textwidth]{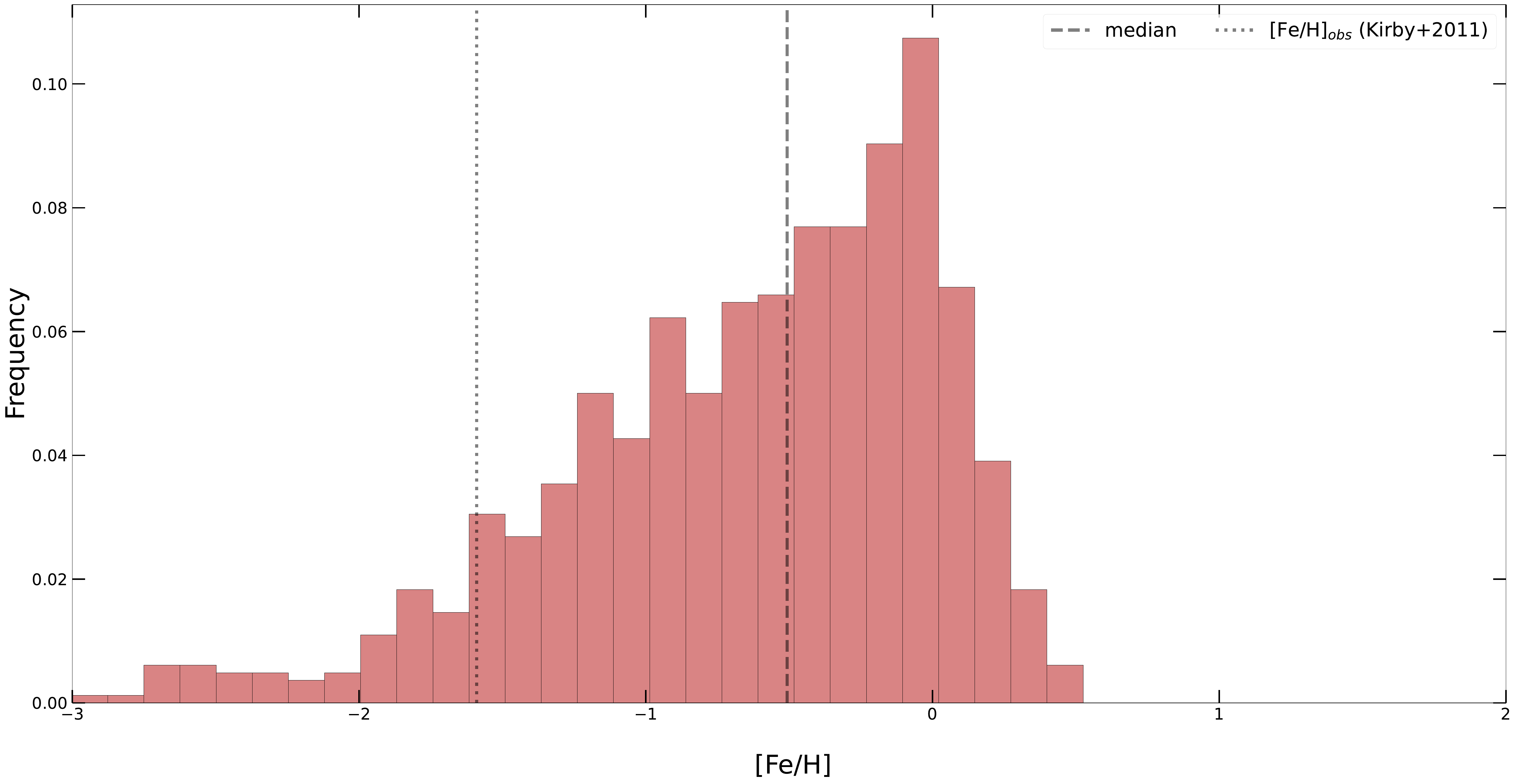}
    \vspace*{2mm}
    \caption{Distribution of stellar abundance ratios [Fe/H] for the fiducial simulation of Leo II ($\eta60$vCalc) at 13.7 Gyr.} 
    \label{fig:FeH}
\end{figure*}

The stellar metallicity distribution regarding the abundance ratios [Fe/H] is depicted in Fig.~\ref{fig:FeH}. Surprisingly, there are stars with supersolar values of [Fe/H]. Moreover, the median value for this distribution at the end of the simulation was calculated as -0.51 dex, which is $\sim1$ dex higher than the value of $-1.59$ assumed as the median for stars in Leo II \citep{Kirby2011multi}, and $\sim0.6$ dex higher than $-1.13^{+0.09}_{-0.31}$ estimated by \citet{Dolphin2002}. The asymmetric nature of the distribution, with the presence of a metal-poor tail, indicates the pristine nature of the gas at the beginning of the simulation. Apart from the existence of this gas, with the potential to dilute the interstellar medium throughout the star formation history, the existing tension with the stellar metallicity constraint for Leo II may suggest alternative gas infall dynamics not accounted for in this isolated simulation.

Following \citet{Lanfranchi&Matteucci2010} discussion, it could be argued that the presence of a peak in [Fe/H] distribution slightly below solar values in Fig.~\ref{fig:FeH} suggest either an excessively high star formation rate or an extended star formation history. However, when we compare both the SFH derived from the chemical evolution models and the results from the SPH simulations, we find that the peak for the star formation rate here is located at $\sim$ 1 dex lower (Fig.~\ref{fig:sfr}). Therefore, it would be reasonable to expect also a lower stellar metallicity, which is not the case. On the other hand, the duration of star formation does not pose an issue, as the majority of stars are formed within the first 7 Gyr in both models. 

\begin{figure*}[h]
    \includegraphics[width=\textwidth]{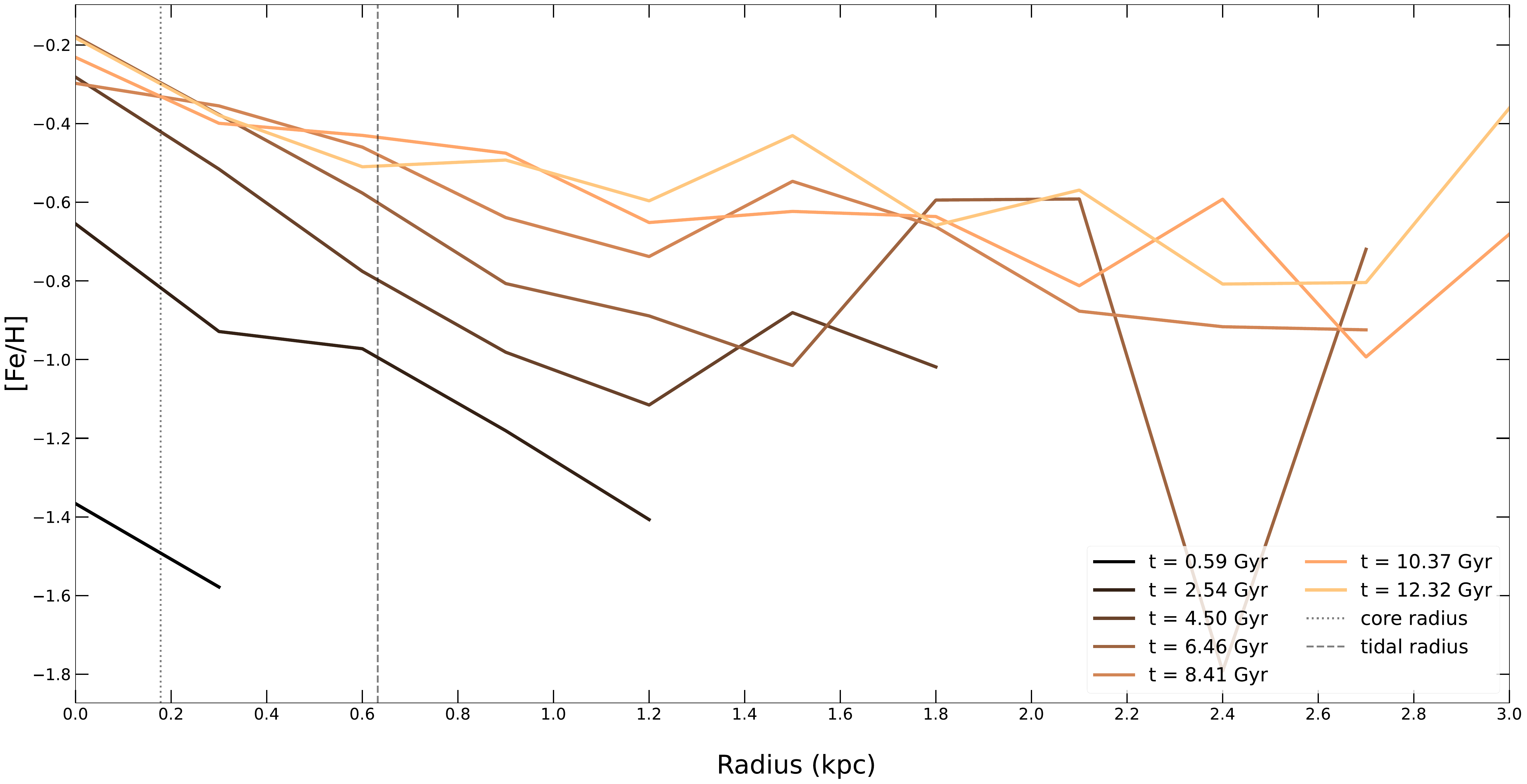}
    \vspace*{2mm}
    \caption{Radial distribution of [Fe/H] in stars for the fiducial simulation of Leo II ($\eta60$vCalc) at 13.7 Gyr.}
    \label{fig:iron_prof}
\end{figure*}

Fig.~\ref{fig:iron_prof} displays the radial profiles of [Fe/H] in stars over time for simulation $\eta60$vCalc. The general trend observed is the presence of abundance gradients, typically lower than 1 dex. These gradients are most pronounced within the central kiloparsec and exhibit some oscillations at larger radii. Such oscillations may be attributed to limited statistics, as there are fewer star particles in these regions. Additionally, the progressive enrichment with iron over time is evident through the vertical shifts of the curves towards higher iron abundances. This enrichment, when considered alongside radial profiles of [Fe/H] in gas, rules out the possibility that more pristine gas from the reservoir was falling to the central regions for further star formation in the simulations. It could prevent the continuation of the fast enrichment of the ISM in iron and, consequently, the enrichment of iron in stars above the observational constraints available for Leo II (further discussion in section \ref{metallicity}.)

Iron is primarily deposited into the interstellar medium by supernovae Ia. The fact that its abundance in the simulation exceeds the observational constraint, $\text{[Fe/H]}_{\text{obs, LeoII}} = -1.59$ \citep{Kirby2011multi}, suggests some possible scenarios, which can be summarized as follows:

(I) A final stellar mass that is excessively high, resulting in a greater injection of iron into the interstellar medium than expected, thereby leading to a higher metallicity.

(II) A star formation history that is overly long, characterized by gas that progressively accumulated iron from Type Ia supernovae over time, without efficient dilution by more pristine gas.

(III) Inefficient dilution of elements during the star formation history, causing faster and excessive iron enrichment over time. 

The first conjecture can be ruled out by comparing the final stellar mass in the fiducial SPH simulation ($\sim2 \times 10^6$ M$_{\odot}$) of Leo II with the stellar mass employed by \citet{Lanfranchi&Matteucci2010} to properly reproduce the metallicity constraints for the galaxy. Their model of dSph galaxies starts with a continuous infall of pristine gas with mass $\sim10^8$ M$_{\odot}$ and achieve a final stellar mass higher than $10^6$ M$_{\odot}$. Furthermore, \citet{Woo2008} in their study of scaling relations for Local Group dwarf galaxies, also estimated the stellar mass of Leo II to be $\sim10^6$ M$_{\odot}$.

Against the second hypothesis there is compelling evidence that, despite more recent episodes of star formation, the majority of stars in Leo II formed during a primary period lasting less than 7 Gyr, which was the estimated in the chemical modelling of this galaxy in \citet{Kirby2011metals}. \citet{Dolphin2002} independently estimated the star formation duration of Leo II as 9.4 Gyr (starting $\sim15$ Gyr ago) by applying numerical methods to derive the SFH from color-magnitude diagrams in seven local dwarf galaxies. The SFH of our simulated galaxy (Fig.~\ref{fig:sfr}) agrees with their estimated star formation duration, with only a minor episode of star formation in the last 4 Gyr of galactic evolution.  

Regarding the third conjecture, chemical modeling studies of Leo II by \citet{Lanfranchi&Matteucci2010} successfully replicated the metallicity constraints by adjusting various parameters, including the gas infall timescale. Additionally, their model assumes instantaneous metal mixing in the interstellar medium following ejection from stars. While this assumption can be considered as an upper limit for the actual gas dilution dynamics in Leo II, it suggests that in our SPH simulations the gas infall regime may have been too rapid from the outset of galactic evolution. A delayed and/or intermittent infall regime (leading to a distinctive SFH profile), coupled with a greater influx of pristine gas, could provide the appropriate dynamics to prevent excessive metal enrichment in the central region, where the majority of star formation occurs in the simulated galaxy. We further investigated this assumption by indirectly increasing the gas dynamical timescale ($t_{\text{dyn}} \sim1 / \sqrt{G\overline{\rho}}$) in a simulation discussed in section \ref{gas-reservoir}, where we reduced the baryon fraction. 

While analyzing their data for Leo II, \citet{Koch2007} did not find conclusive indications of any radial metallicity gradient for stars. However, the gradients observed in our fiducial simulation are consistent with the findings of \citet{komiyama2007}, who provided evidence of more metal-poor and older stars in the galactic outskirts (also corroborating the results depicted in Fig.~\ref{fig:stellar-age-profile}). They reached these conclusions through an analysis of a morphological index for HB stars, extending beyond the tidal radius of Leo II. Furthermore, the gradients we observe in our simulation are lower than the value of -0.18 dex per core radius found by \citet{Kirby2011multi}.  

The tension observed in replicating the metallicity constraints, both in our fiducial simulation and in other simulations considered in Table~\ref{tab:simul_table} carries broader implications, which we delve into in sections \ref{metallicity} and \ref{discussion}. This result, coupled with the presence of a super-solar metallicity tail evident in Fig.~\ref{fig:iron_prof}, raises the possibility that the dilution of metals within the interstellar medium remains inefficient in these isolated simulations. One potential avenue to alleviate this tension would involve direct infalls of more pristine gas over time, as also advocated by \cite{Koch2007} and \cite{Kirby2011metals} for Leo II. However, an isolated simulation could only reproduce it with ad hoc procedures. For this purpose, upcoming cosmological simulations will be carried out to address this issue self-consistently.

\subsection{Thermal-only stellar feedback}

\begin{figure*}
    \includegraphics[width=\textwidth]{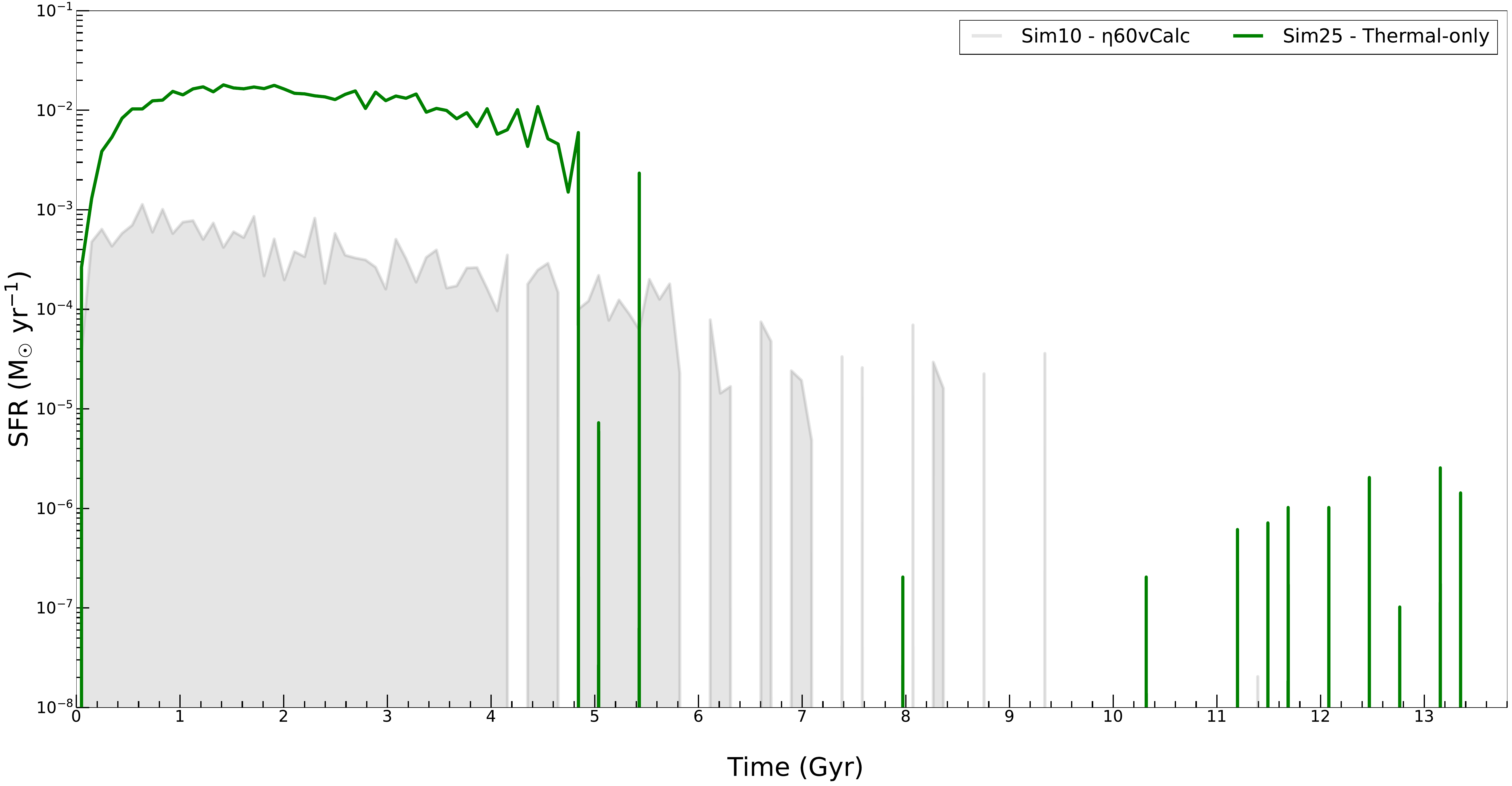}
    \vspace*{2mm}
    \caption{Star formation history for the simulation with thermal-only stellar feedback, compared with the fiducial simulation ($\eta60$vCalc).}
    \label{fig:sfh-thermal}
\end{figure*}

We conducted a simulation using thermal-only stellar feedback, employing the same parameters of our fiducial case (section \ref{optsim}). Table~\ref{tab:thermal} presents a comparison between these simulations, highlighting selected constraints. It is evident that both simulations resulted in gas depletion in the tidal region of the simulated galaxy, but with striking differences in their evolutionary paths. The most significant discrepancy lies in the final stellar mass produced in this scenario, which exceeded that of the simulation with kinetic+thermal feedback by more than 1 dex. The SFH of the galaxy with thermal-only feedback is illustrated in Fig.~\ref{fig:sfh-thermal}. Notably, it reveals a continuous star formation episode spanning $\sim4.8$ Gyr, interspersed with episodic star formation events at lower rates until recent times. The peak of star formation occurs between 1 and 2 Gyr at a rate $\sim10^{-2}$ M$_{\odot}$ yr$^{-1}$, a magnitude roughly 1 dex higher than that observed in our fiducial simulation.

\begin{table}[h]
	\centering
        \caption{Influence of the type of stellar feedback in the simulations.}
        \label{tab:thermal}
        \begin{tabular}{cccc} 
		\hline
	    Simulation & $\eta60$vCalc & thermal-only\\
		\hline
		Residual tidal gas mass (M$_{\odot}$) & 0 & 0\\
	  x$_{\text{gas, outflow}}$ at t = 12 Gyr & 0.67 & 0.67\\
		Stellar mass (M$_{\odot}$) & $2.0 \times 10^6$ & $5.1 \times 10^7$\\
		$[\text{Fe/H}]_{\text{median}}$ in stars & -0.51 & 0.40\\
        Median stellar age (Gyr) & 11.8 & 11.7\\
        \hline
        \end{tabular}
\end{table}

\begin{figure*}
    \includegraphics[width=\textwidth]{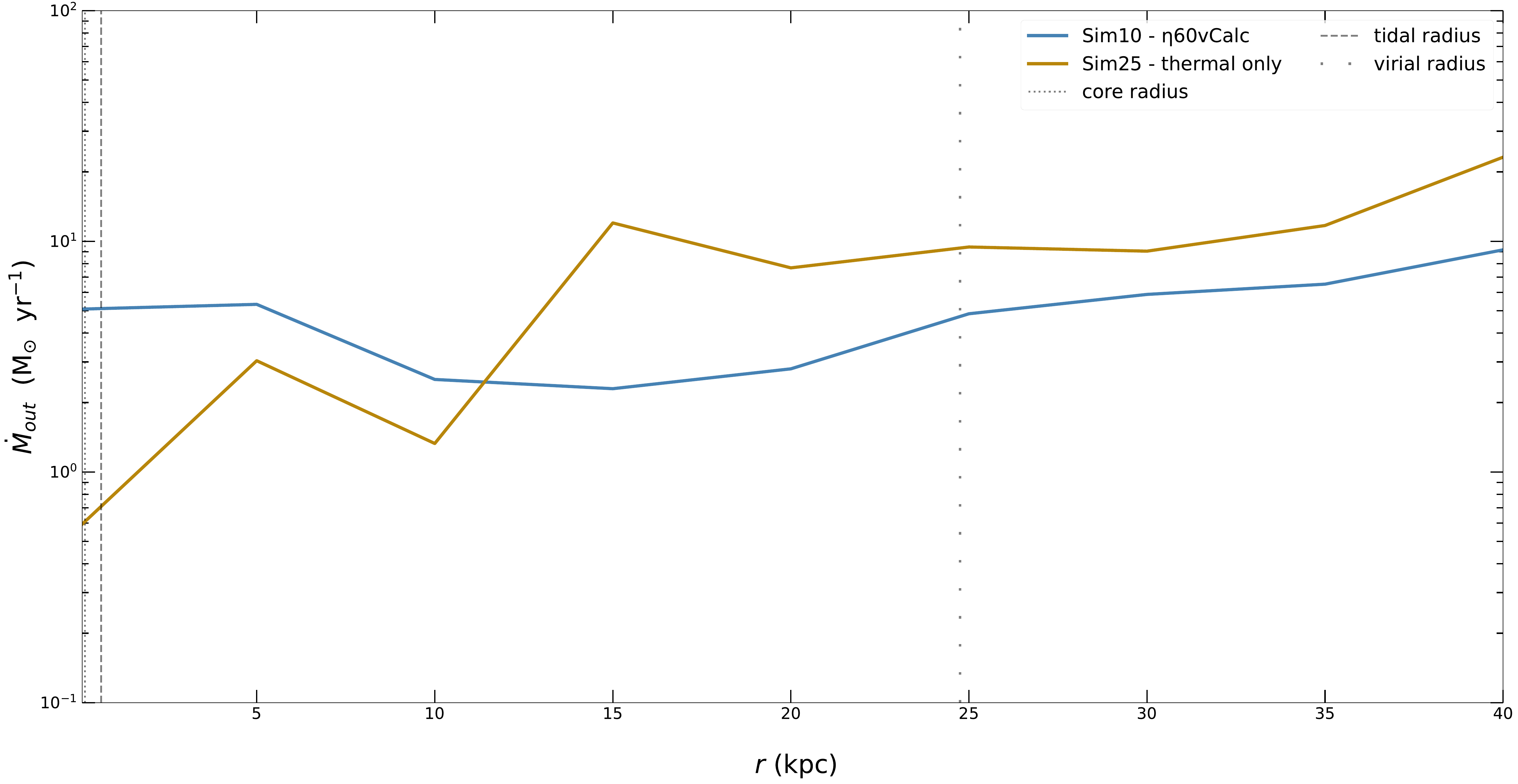}
    \vspace*{2mm}
    \caption{Radial profile of gas outflowing rates ($v_{\text{rad}}  > v_{\text{esc}}$) for thermal-only stellar feedback and comparison with the fiducial simulation ($\eta60$vCalc) at 6.85 Gyr.}
    \label{fig:Mout-thermal}
\end{figure*}

Interestingly, the mass fraction of outflowing gas at 12 Gyr was not changed by removing the kinetic component of the stellar feedback (Table~\ref{tab:thermal}). This seemingly unexpected result can be explained by the significant increase in the number of stars generated in this simulation (more than 1 dex in additional stellar mass), which counterbalances the less intense feedback in the thermal-only scenario. In fact, this trend can also be seen in Fig.~\ref{fig:Mout-thermal}, which shows a snapshot of the local mass outflow rate as function of the galactic radius. At the time considered in this snapshot, the continuous and strongest period of star formation has already ended in both simulations (see section~\ref{optsim}). Fig.~\ref{fig:Mout-thermal} illustrates that the outflowing rates are higher for the combination of kinetic and thermal feedback up to $\sim$ 14 kpc. However, beyond this point, the thermal-only stellar feedback generates higher rates. 

The production of more than 1 dex of additional stellar mass in the thermal simulation also resulted in a greater mismatch in the stellar metallicity constraint, increasing the value of [Fe/H] in 0.91 dex. The evolution of the galaxy under such conditions generate super-solar stars, whose metallicities differ by $\sim$2 dex from the observed value of [Fe/H]$_{\text{obs}} = -1.59$ in \citet{Kirby2011multi}.

\subsection{Gas Depletion}

In this section, we investigate the influence of the parameter space of the kinetic stellar feedback comprising different combinations of $\eta$ $\times$ $v_{\textit{wind}}$ for Eq.~\ref{eqn:param_space}. Thermal stellar feedback is also implemented in all the simulations. 

\begin{figure*}[hb]
    \includegraphics[width=\textwidth]{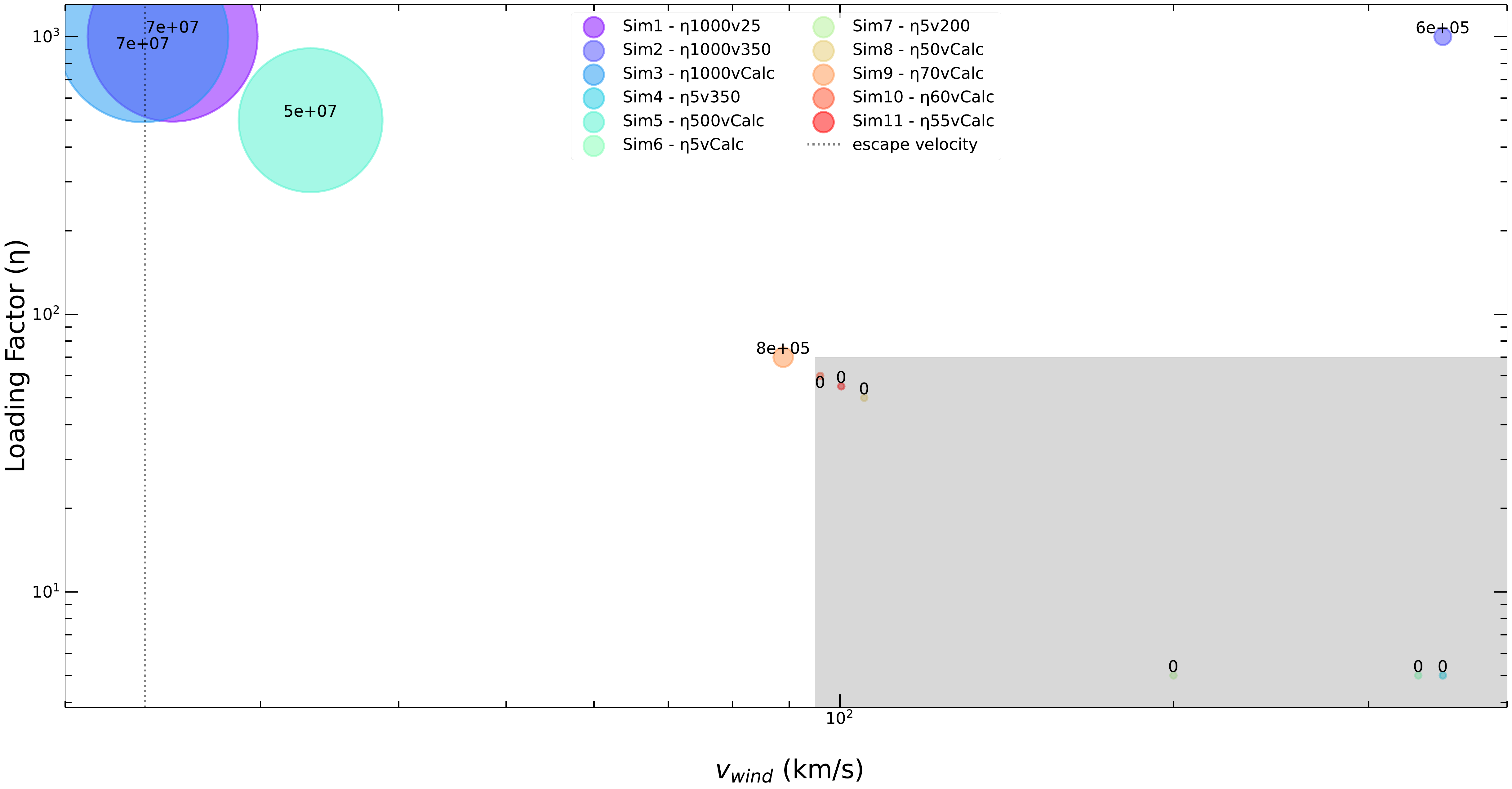}
    \vspace*{2mm}
    \caption{Gas mass (\textbf{\color{blue} M$_{\odot}$}) within the tidal radius of simulated galaxies at the end of simulations (t $\sim$ 13.7 Gyr).}
    \label{fig:tidal_mass}
\end{figure*}

One of the main constraints to be reproduced in simulations of dwarf spheroidal galaxies is the residual gas mass, which must be $\lesssim10^4$ M$_{\odot}$ in the case of Leo II \citep{Grcevich2009}. The low gas content justify the quenching of such galaxies as observed in recent times. Fig.~\ref{fig:tidal_mass} shows the residual gas mass within the tidal radius ($r_t \sim650$ pc) for various combinations of mass loading factors and input wind velocities in the stellar feedback model.

Fig.~\ref{fig:tidal_mass} reveals that the residual gas mass within the tidal region spans seven orders of magnitude, depending on the choice of the model parameters. When low values of wind velocity are combined with high mass loading factors, galaxies with excessive gas mass are generated. It may be associated to gas that is expelled during star formation and evolution but, due to loss of momentum by shocks and thermalization with other gas particles during its path outwards, it is able to return to the central regions of the galaxy at later times. Notably this trend persists even in simulations with wind injection velocity higher than the escape velocity ($v_{\textit{esc}} \sim24$ km s$^{-1}$), such as for $\eta500$vCalc ($v_{\textit{wind}} \sim33$ km s$^{-1}$). This can be attributed to the fact that the wind velocity is injected into gas particles within the central regions, mainly in the central 3 kpc. Given that the virial radius of the simulated galaxies is $r_{200} \sim25 $ kpc, wind particles have a significant probability of encountering and colliding with other gas particles before escaping the galactic gravitational pull. This process results in momentum transfer and, eventually, the return of gas to the galactic center. 

By examining the parameter space depicted in Fig.~\ref{fig:tidal_mass}, it becomes apparent that mass loading factors higher than 100 can be discarded. The optimal parameter combinations that resulted in no remaining gas at the end of simulations are shaded in gray in this figure. Essentially, these correspond to mass loading factors $\eta \lesssim 70$ and wind injection velocities in the range $90 \lesssim v_{\text{wind}} \lesssim 350$ km s$^{-1}$.

\begin{figure*}[hb]
    \includegraphics[width=\textwidth]{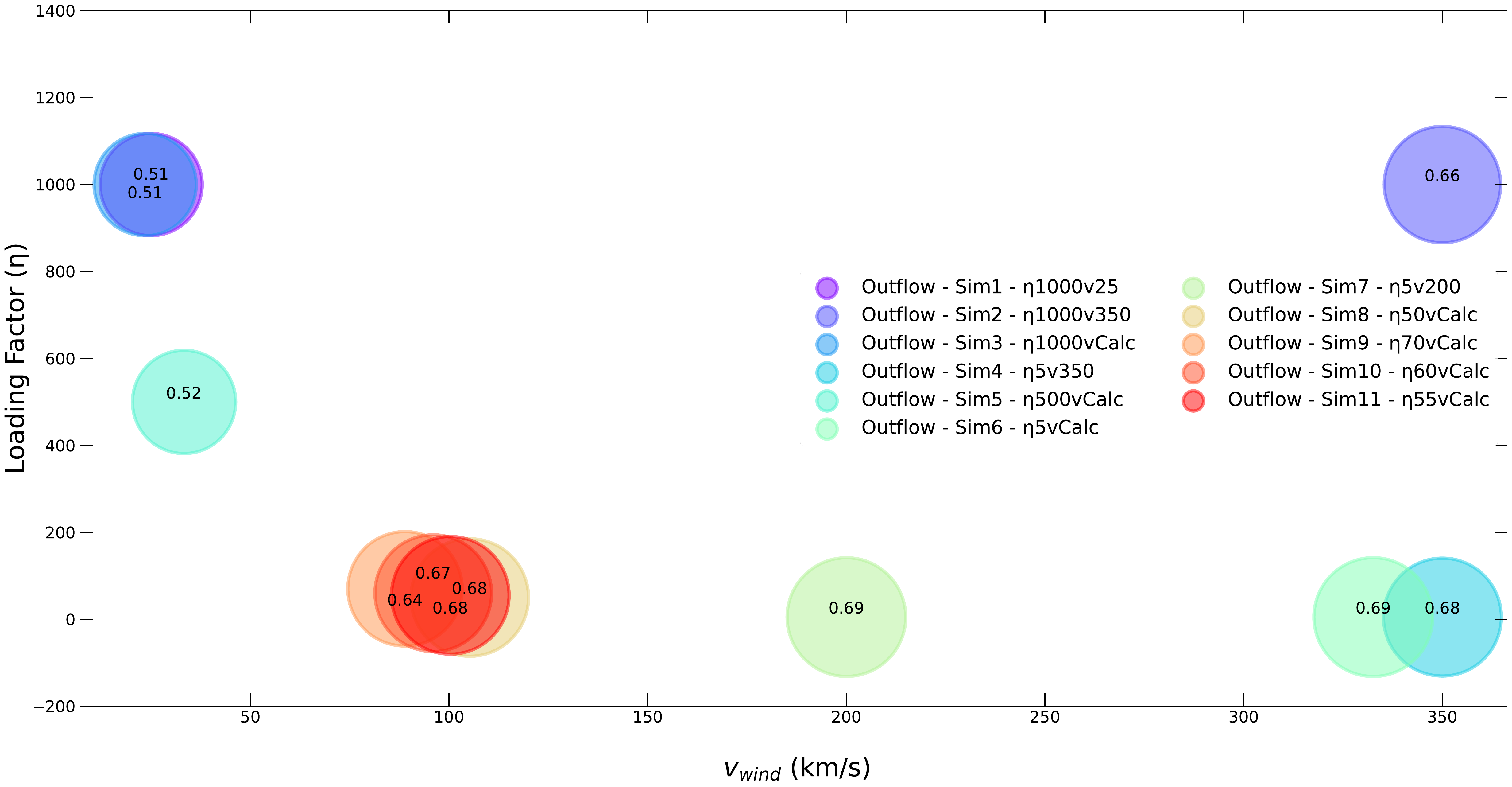}
    \vspace*{2mm}
    \caption{Total mass fraction of gas classified as outflow projected against the parameter space for the stellar feedback tests at a common reference time of t = 12 Gyr (when the SF has ceased in most cases).}
    \label{fig:outflow_frac}
\end{figure*}

Fig.~\ref{fig:outflow_frac} shows the total mass fraction of gas classified as outflow using the criterion of $v > 0$ at any $r > R_{200}$ for all the simulations listed in Table~\ref{tab:simul_table}. The chosen time for comparison among the simulations is $t = 12$ Gyr. Simulations with the lowest mass loading factors $\eta \lesssim 100$ yielded the highest absolute values of outflow mass fraction ($x_{\text{out}} \sim0.64-0.69$), with exception for the simulation $\eta1000v350$, which produced $\eta \sim0.66.$ The $\chi$ value for this simulation is $\chi = 111$, corresponding to an injection of energy two orders of magnitude greater than the available energy from supernovae on average. However, this additional energy did not lead to a significant increase in gas depletion within the galaxy, as $\sim6\times10^5$ M$_{\odot}$ remained in the tidal region by the end of simulation (Fig.~\ref{fig:tidal_mass}). 

As a general result, in all the simulations examined, most of the gas mass was expelled into the intergalactic medium, leaving only $\sim30\%$ of the gas either inflowing or lacking sufficient kinetic energy to overcome the galaxy's gravitational potential. This remaining gas has the potential to refuel star formation in later epochs. 

In simulations that produced the highest amount of gas mass within the tidal radius, characterized by high mass loading factors and low wind ejection velocities (Fig.~\ref{fig:tidal_mass}), approximately $51\%$ of the gas particles would be related to outflows in Fig.~\ref{fig:outflow_frac}. The discrepancy between the depleted-gas galaxies and gas-rich ones is around $17\%$. The total gas reservoir in these simulations is $\sim3.2 \times 10^8$ M$_{\odot}$ (see Table~\ref{tab:config_table}). Note that $17\%$ of this initial gas reservoir is $\sim10^7$ M$_{\odot}$, which is the same order of magnitude of the gas found within the tidal regions of gas-rich galaxies (Fig.~\ref{fig:tidal_mass}). Consequently, it can be inferred that the gas amount difference among these simulations is not located in the outskirts, but rather in the central regions of the galaxy and prone to star formation once the appropriate conditions are met. 

The simulated galaxies which retained considerable amount of gas by the end of the simulation exhibit characteristics more akin to dwarf irregular galaxies (dIrrs) than dwarf spheroidals. Some of these galaxies even display signs of gas rotation, indicative of a disk-like structure. According to \citet{Mayer2006}, dIrrs could potentially transform into gas-depleted dSphs through interactions with the host galaxy and internally-driven outflows resulting from supernovae. For such systems, stellar feedback alone may prove insufficient to remove gas and subsequently quench the galaxy. In this case, the transition to a dSph state would likely necessitate additional influences, such as tidal interactions and ram-pressure effects from interactions with other galaxies.

\subsection{Stellar metallicity} \label{metallicity}

As already noted by \citet{Collins&Read2022}, for nearby galaxies with measured stellar metallicities, there is often no comparable gas metallicity data available. This is primarily because most local dwarf galaxies are quenched systems that lack significant amounts of gas. In this way, the stellar metallicity constraint remains the unique available tracer for assessing the total metal budget in dwarf spheroidal galaxies. \citet{Kirby2011metals} observed that the metal content of stars in dSphs was significantly lower than expected by closed-box models of chemical evolution. The authors argued that gas outflows could remove most of the metals produced by such galaxies, with dSphs potentially losing over $96\%$ of the metal content produced by stars. Table \ref{tab:metals} present the results for simulations that reproduced the most plausible values for the stellar metallicity (sub-solar values) in terms of [Fe/H].  

\begin{table*}[h]
	\centering
        \caption{Optimal SPH simulations for stellar metallicity constraints.}
        \label{tab:metals}
        \resizebox{\textwidth}{!}{\begin{tabular}{ccccccc} 
		\hline
	    Simulation & IMF & Residual tidal gas mass (M$_{\odot}$) & x$_{\text{gas, outflow}}$ at t = 12 Gyr & Stellar mass (M$_{\odot}$) & $[\text{Fe/H}]_{\text{median}}$ & Median stellar age (Gyr)\\
		\hline
        $\eta$5vCalc & Chabrier & 0 & 0.69	& $9.5 \times 10^5$ & $-0.26$ & 12.9\\
        $\eta$50vCalc & Chabrier & 0 & 0.68 & $1.3 \times 10^6$ & $-0.74$ & 10.8\\
        $\eta$60vCalc & Chabrier & 0 & 0.67 & $2.0 \times 10^6$ & $-0.51$ & 11.8\\
        $\eta$55vCalc & Chabrier & 0 & 0.68 & $1.5 \times 10^6$ & $-0.65$ & 12.1\\
        $\eta$60vCalcBF4 & Chabrier & 0 & 0.67 & $2.2 \times 10^6$ & $-0.14$ & 11.7\\
        $\eta$60v117BF4 & Chabrier & 0 & 0.65 & $7.0 \times 10^5$ & $-0.62$ & 12.8\\
        $\eta$60v117BF4-Sal & Salpeter & 0 & 0.73 & $6.4 \times 10^5$ & $-0.71$ & 12.8\\
        $\eta$45v135BF4-Sal & Salpeter & 0 & 0.68 & $6.4 \times 10^5$ & $-0.81$ & 12.9\\
        $\eta$60vCalc-b5 & Chabrier & $7.5 \times 10^5$ & 0.67 & $1.1 \times 10^6$ & $-1.07$ & 12.6\\
        $\eta$60vCalc-b30 & Chabrier & 0 & 0.67 & $3.8 \times 10^6$ & $-0.69$ & 12.1\\
        \hline
        \end{tabular}}
\end{table*}

In the isolated galactic simulations conducted in this study, the most challenging constraint to replicate was the metallicity of stars. While the models developed are not strictly closed-box models, they do incorporate certain assumptions about the initial gas distribution. Furthermore, no additional ad-hoc mechanisms were introduced to modify how gas would supply the regions conducive to star formation over time. It's worth noting that in these simulations, the galaxy remains stationary, and no other environmental effects were considered that could potentially alter the inflow and outflow dynamics of gas. These mechanisms will be explored in future work, with a focus on cosmological simulations.

\begin{figure*}
    \includegraphics[width=\textwidth]{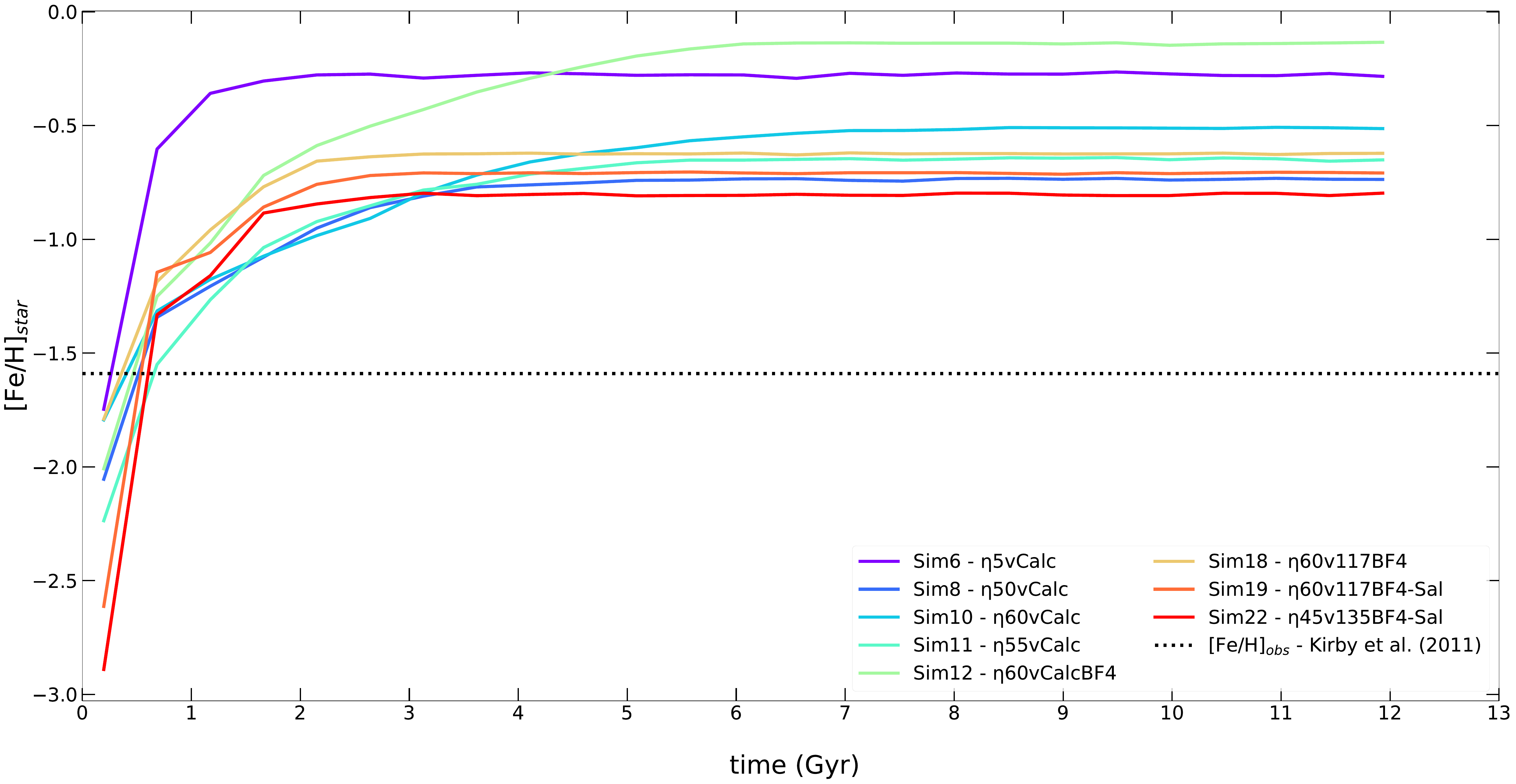}
    \vspace*{2mm}
    \caption{Evolution of [Fe/H] in stars for simulations that produced the lowest values (sub-solar) of stellar metallicity.}
    \label{fig:metal-evol}
\end{figure*}

As shown in Fig.~\ref{fig:metal-evol}, none of the simulations succeeded in reproducing the reference value of [Fe/H]$_{\text{obs}}$ = $-1.59$ \citep{Kirby2011multi}. Furthermore, it was verified that the reference value for [Fe/H] is attained before 1 Gyr in all simulations. The hypothesis that the SF in Leo II could have been as brief as 1 Gyr can be dismissed by analysis of color-magnitude diagrams for this dwarf galaxy, as previously discussed by \citet{Kirby2011metals} in their work on metals in Leo II. 

Even the most successful simulation in terms of reproducing the stellar metallicity could not capture the abrupt decline in the frequency of metal-rich stars, as observed in \citet{Kirby2011metals}. Instead, the simulations exhibit a more dispersed distribution around the median, as illustrated in the example shown in Fig \ref{fig:FeH}, for our fiducial simulation.

The following subsections explore in more detail changes that were implemented in the simulations aiming to improve the replication of the stellar metallicity constraints for Leo II. 

\subsubsection{Binary fraction}

In most of the simulations, a binary fraction (BF) of 0.1 was employed for the purpose of comparing results with other simulations that use this fiducial value \citep[e.g.][]{Barai2015}. However, evidence in the literature suggests different binary fractions for dwarfs \citep[e.g.][]{Spencer2018}. In the specific case of Leo II, \citet{Spencer2017binary} found values ranging from 0.30 to 0.34. Hence, the simulations 12-22 were performed with an upper limit of 0.4 for the binary fraction to examine its influence. Table \ref{tab:binary_table} provides a comparison of our fiducial and two analogous simulations that differ solely in their choice of binary fraction. 

\begin{table}[h]
	\centering
        \caption{Influence of the binary fraction in the simulations.}
        \label{tab:binary_table}
        \begin{tabular}{cccc} 
		\hline
	    Simulation & $\eta60$vCalc & $\eta60$vCalcBF4 & $\eta60$vCalcBF4-v2\\
		\hline
		Residual tidal gas mass (M$_{\odot}$) & 0 & 0 & 0\\
		x$_{\text{gas, outflow}}$ \text{at t = 12 Gyr} & 0.67 & 0.67 & 0.66\\
		Stellar mass (M$_{\odot}$) & $2.0 \times 10^6$ & $2.2 \times 10^6$ & $2.2 \times 10^6$\\
		$[\text{Fe/H}]_{\text{median}}$ & -0.51 & -0.14 & -0.16\\
        \hline
        \end{tabular}
\end{table}

\begin{figure*}
    \includegraphics[width=\textwidth]{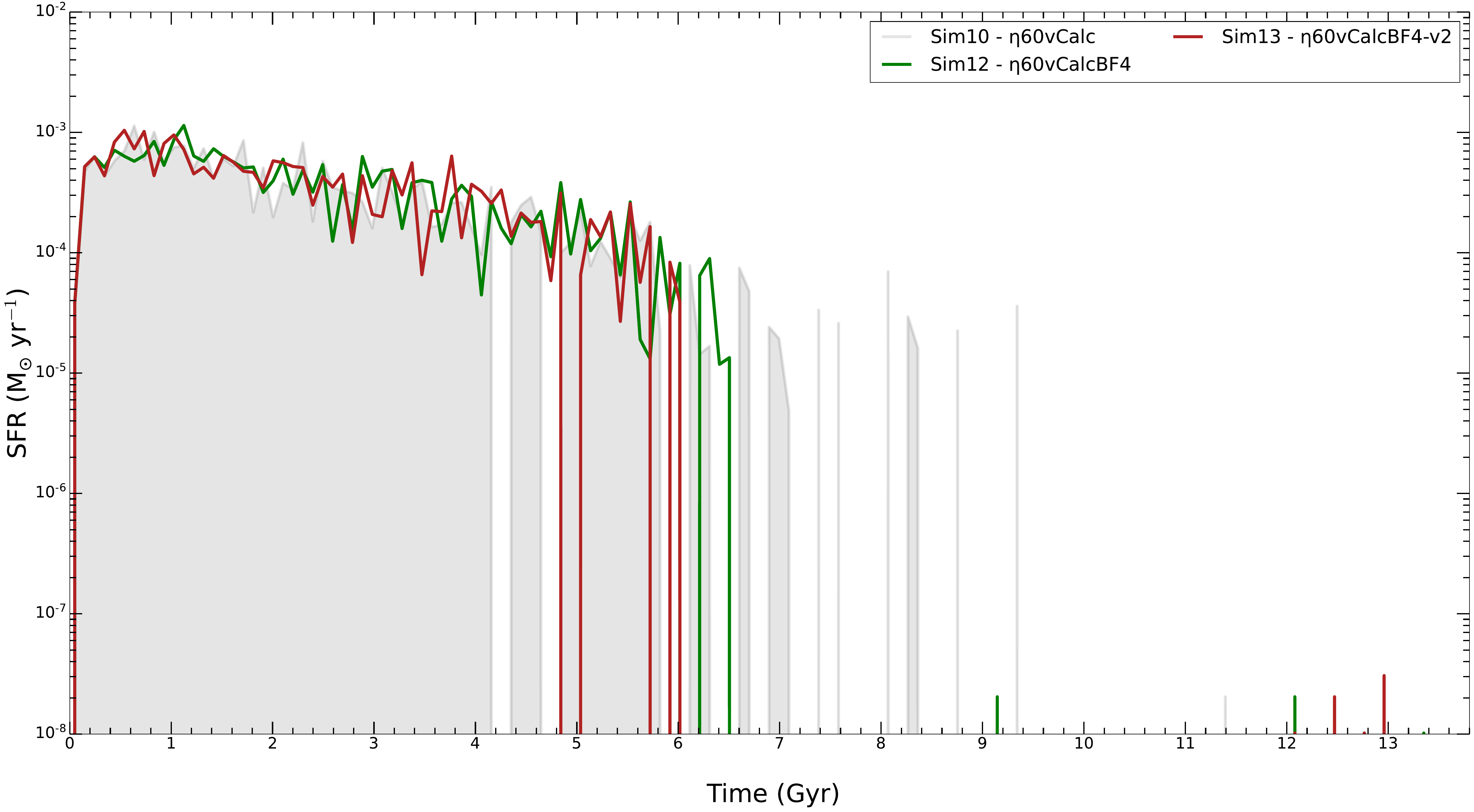}
    \vspace*{2mm}
    \caption{Star formation histories comparing the fiducial simulation ($\eta60$vCalc) with BF = 0.1 and two other simulations with BF = 0.4.}
    \label{fig:sfr-binary}
\end{figure*}

Table \ref{tab:binary_table} indicates that the gas depletion was not affected by changing the binary fraction from 0.1 to 0.4, as no gas particles were found within the tidal region at the end of all the simulations. Regarding the mass fraction of outflowing gas particles, no significant differences were observed. However, the final stellar mass was increased by $\sim$ 0.2 dex by increasing the BF. To investigate potential distinctions in the buildup of stellar mass over time, the SFHs of the three simulations were compared in Fig.~\ref{fig:sfr-binary}.

The trends observed in the star formation rates in Fig.~\ref{fig:sfr-binary} remain virtually the same during the initial 4 Gyr of continuous star formation, both in terms of the peak of these distributions ($\sim10^{-3}$ M$_{\odot}$ yr$^{-1}$) and the continuous initial decline of the curve. However, striking differences were observed in the number of star formation episodes, which decreased when the BF was increased. Note, however, that they were not substantial enough to significantly change the final stellar mass, as indicated in Tab.~\ref{tab:binary_table}. 

The most significant differences observed when BF was changed were the values of the median stellar metallicities in the simulated galaxies, as illustrated in Tab.~\ref{tab:binary_table}. Notably, there was an increase of $\sim0.36$ dex for BF = 0.4. This difference can be attributed to the higher number of binary systems where supernova type Ia explodes, which is the main source of iron in the universe. Inflows of pristine gas and the effects of tidal and ram-pressure from a host galaxy could potentially counteract this enrichment trend. However, as all the simulations in this study were conducted in isolation to mimic the predicted lack of influence from a host galaxy on the evolution of Leo II, such effects were not considered. Furthermore, the higher mismatch in the reproduction of median stellar metallicity for Leo II indicates that the tuning of the stellar feedback parameters is dependent on the choice of binary fraction.

\subsubsection{Initial Mass Function}

To evaluate the impact of the IMF, some of the simulations were run switching it from Chabrier to Salpeter. Tab.~\ref{tab:IMF_table} presents key variables of interest for both cases. The other parameters regarding the galaxy initial conditions and the stellar feedback were kept fixed to isolate the effects of the IMF in the results. 

\begin{table}[ht]
	\centering
        \caption{Influence of the initial mass function in the simulations.}
        \label{tab:IMF_table}
        \begin{tabular}{ccc} 
		\hline
	    Simulation & $\eta60$vCalcBF4 & $\eta60$vCalcBF4-Sal\\
		\hline
		Initial Mass Function & Chabrier & Salpeter\\
        Residual tidal gas mass (M$_{\odot}$) & 0 & $1.6 \times 10^6$\\
		x$_{\text{gas, outflow}}$ \text{at t = 12 Gyr} & 0.66 & 0.60\\
		Stellar mass (M$_*$/M$_{\odot}$) & $2.2 \times 10^6$ & $1.4 \times 10^7$\\
		$[\text{Fe/H}]_{\text{median}}$ & -0.16 & 0.51\\
        Median stellar age (Gyr) & 11.7 & 7.8\\
        Wind velocity (km/s) & 95.6 & 75.5\\
        Mass fraction of stars in range (8 - 100) M$_{\odot}$ & 0.229 & 0.139\\
        Feedback energy per formed solar mass in stars (erg) & $1.1 \times 10^{49}$ & $6.8 \times 10^{48}$\\
        SNe II & 5232 & 13750\\
        SNe Ia & 970 & 1988\\
        Ratio SNe Ia/II & 0.19 & 0.14\\
        SNe II/M$_*$ & $2.4 \times 10^{-3}$ & $9.8 \times 10^{-4}$\\
        SNe Ia/M$_*$ & $4.4 \times 10^{-4}$ & $1.4 \times 10^{-4}$\\
        \hline
        \end{tabular}
\end{table}

Significant differences are evident in all the variables analyzed in Tab.~\ref{tab:IMF_table}. The shift to the Salpeter IMF led to less efficient gas depletion within the tidal radius, for the same calibration of stellar feedback. The presence of $\sim10^6$ M$_{\odot}$ of gas in the galaxy is two orders of magnitude greater than the upper limit identified by \citet{Grcevich2009}. It is also reflected in a reduction of 6$\%$ in the mass fraction of outflowing gas particles at the end of simulation. The higher availability of gas further led to a higher final stellar mass, with a difference of $\sim$ 0.9 dex. Consequently, this more intense star formation produced stars with higher metallicities, reaching super-solar levels. 

\begin{figure*}
    \includegraphics[width=\textwidth]{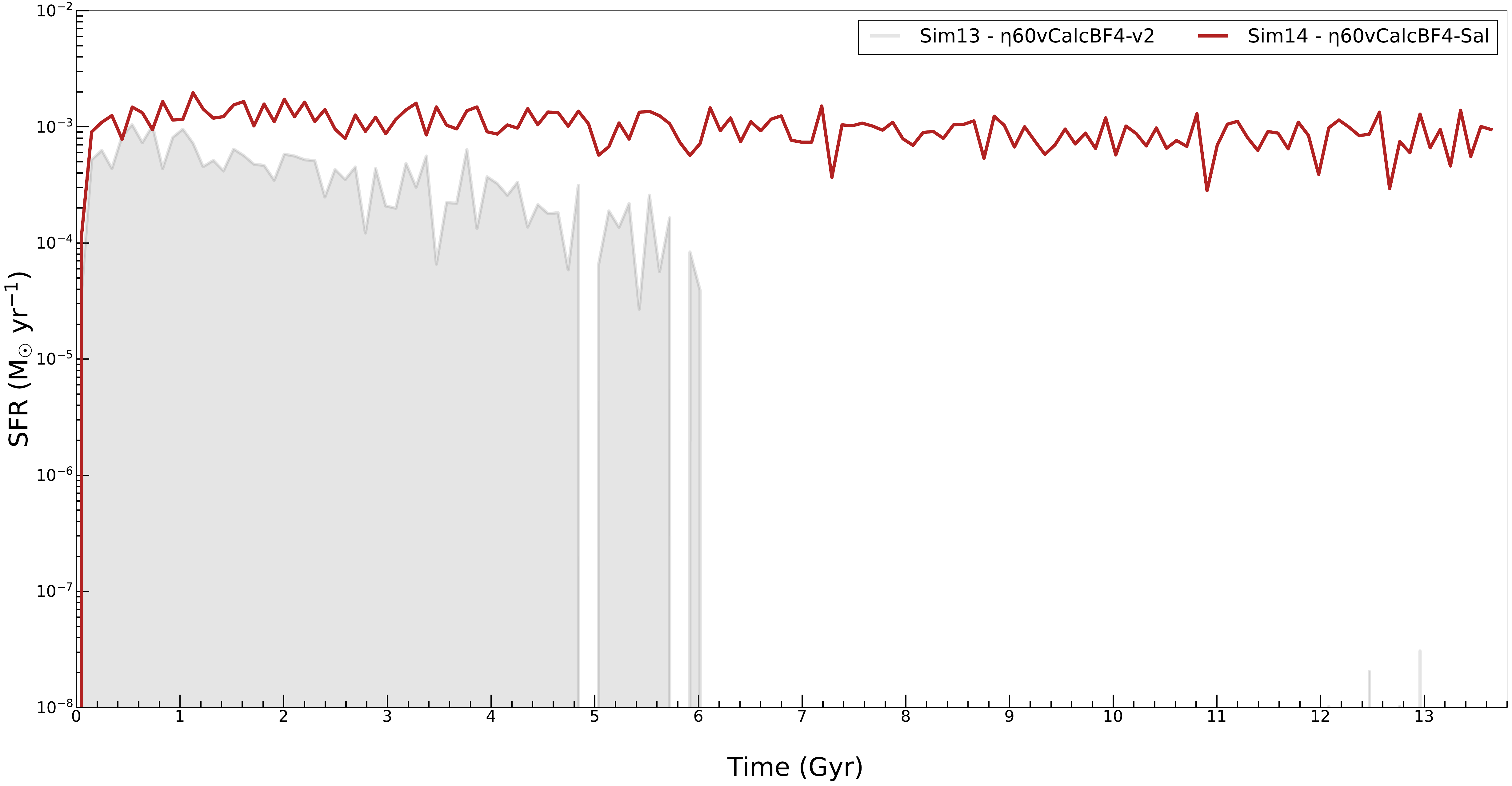}
    \vspace*{2mm}
    \caption{Star formation histories for a simulation with Chabrier ($\eta60$vCalcBF4-v2) and other with a Salpeter IMF ($\eta60$vCalcBF4-Sal).}
    \label{fig:sfr-IMF}
\end{figure*}

To assess the influence on the SFH, Fig.~\ref{fig:sfr-IMF} displays the SFR for the simulation using the Chabrier IMF ($\eta60$vCalcBF4-v2) and the one using the Salpeter ($\eta60$vCalcBF4-Sal). Changing the IMF to Salpeter resulted in striking differences in the SFH. The simulation with the Chabrier IMF produced $\sim$ 6 Gyr of star formation, while using the Salpeter IMF led to a continuous star formation until the present time. This difference can account for the higher stellar mass formed in the galaxy. 

The stronger star formation observed when switching to the Salpeter IMF can be explained by analyzing the energetics associated with each IMF, which is a consequence of their intrinsic stellar mass distribution. As shown in Table \ref{tab:IMF_table}, the Chabrier IMF results in a higher specific energy release per unit mass of stars formed during galactic evolution. The reduction in energy release with the Salpeter IMF is $\sim38\%$, primarily due to the lower fraction of stars with masses exceeding 8 M$_{\odot}$ in this IMF, which are prone to SNe II explosions. This reduction in energy output is also evident when comparing the number of SNe II, which dominates the energetic budget, by stellar mass formed in each simulation. The Salpeter IMF leads to a reduction of $\sim59\%$ in the specific number of SNII explosions. Consequently, it results in lower feedback specific energy and, given the definition of the energetic balance with a constant $\chi$ in Eq.~\ref{eq:eta}, also in a lower wind velocity (75.5 km s$^{-1}$). This contributes to an overall weaker feedback mechanism, resulting in reduced turbulence, less outflows and, consequently, higher stellar mass and sustained star formation up to the present time, as observed in Fig.~\ref{fig:sfr-IMF}.

The overall picture when analyzing the influence of the IMF indicates that the optimal parameter choice to reproduce the observational constraints is also IMF-dependent in the isolated galactic simulations. Other parameter combinations for the Salpeter IMF were explored (see Tab.~\ref{tab:simul_table}). Some of these simulations, such as $\eta$60v117BF4-Sal and $\eta$45v135BF4-Sal achieved acceptable gas depletion and improved stellar metallicity, but at the expense of a lower final stellar mass. Furthermore, no improvements in the stellar ages were observed in these cases.

\subsubsection{Dark matter halo mass}

\citet{Walker2007} refers to a lower value for the DM halo mass ($4 \times 10^8$ M$_{\odot}$) for Leo II when compared to \citet{Strigari2007}. In order to test the effect of such a lower limit, two simulations were performed and some of their key results were compared to those of a fiducial simulation $\eta60$vCalcBF4. The findings are summarized in Table~\ref{tab:lowmass_table}.

\begin{table}[ht]
	\centering
        \caption{Influence of the dark matter halo mass in the simulations.}
        \label{tab:lowmass_table}
        \begin{tabular}{cccc} 
		\hline
	    Simulation & $\eta60$vCalcBF4 & $\eta60$vCalcBF4-LM & $\eta30$vCalcBF4-LM\\
		\hline
		Residual tidal gas mass (M$_{\odot}$) & 0 & $10^4$ & 0\\
		x$_{\text{gas, outflow}}$ at t = 12 Gyr & 0.67 & 1 & 0.68\\
		Stellar mass (M$_{\odot}$) & $2.2 \times 10^6$ & 0 & $6.8 \times 10^4$\\
		$[\text{Fe/H}]_{\text{median}}$ & -0.14 & NA & -0.4\\
        \hline
        \end{tabular}
\end{table}

In an initial simulation where only the DM halo mass was altered to a lower value, the galaxy rapidly dissipated, and star formation became impossible. This resulted in a dark galaxy devoid of baryonic matter. Subsequently, a new set of parameters related to the stellar feedback was tested in simulation $\eta30$vCalcBF4-LM. The final stellar mass formed then was $\sim6.8 \times 10^4$ M$_{\odot}$, nearly 2 dex lower than in simulation $\eta60$vCalcBF4. However, the median stellar metallicity was only slightly sub-solar, and no significant improvements were observed in this key quantity of interest. Consequently, no further simulations with a lighter DM halo were conducted for the purposes of this work.

\subsubsection{Gas reservoir} \label{gas-reservoir}

As previously discussed, the modeling of the infall regime may be the most critical factor affecting the poor reproduction of metallicity and stellar ages for the Leo II galaxy. To investigate this further, we tested different initial conditions for Leo II comprising the initial gas reservoir available for star formation by changing the initial mass fraction of the gas (see Eq.~\ref{eq:gas_ro}). 

Two tests were performed taking the simulation $\eta60$vCalc as the fiducial one ($m_b = 0.16$, representing the mean baryonic mass fraction in the universe), considering also a low-density environment ($m_b = 0.05$) and a high-density one ($m_b = 0.30$). Table~\ref{tab:gas_res} provides a summary of some key results regarding important constraints, compared to a fiducial simulation. Only the initial baryonic mass fraction was varied, while the other parameters remained the same as in simulation $\eta60$vCalc. This procedure, maintaing $b$ constant in Eq.~\ref{eq:gas_ro}, would be equivalent to adjusting the infall timescale ($t_{\text{dyn}} \sim1/ \sqrt{G \overline{\rho}}$), similar to what was done in the chemical models for Leo II by \cite{Lanfranchi&Matteucci2010}. As a general result, the outflowing gas fraction at the end of the simulation did not change when varying the initial baryon mass fraction. 

\begin{table}
	\centering
        \caption{Influence of the initial gas reservoir in the simulations.}
        \label{tab:gas_res}
        \begin{tabular}{cccc} 
		\hline
	    Simulation & $\eta60$vCalc & $\eta60$vCal-b5 & $\eta30$vCal-b30\\
		\hline
		Residual tidal gas mass (M$_{\odot}$) & 0 & $7.5 \times 10^5$ & 0\\
		x$_{\text{baryon},0}$ & 0.16 & 0.05 & 0.30\\
        x$_{\text{gas, outflow}}$ at t = 12 Gyr & 0.67 & 0.67 & 0.67\\
		Stellar mass (M$_{\odot}$) & $2.0 \times 10^6$ & $1.1 \times 10^6$ & $3.8 \times 10^6$\\
		$[\text{Fe/H}]_{\text{median}}$ in stars & -0.51 & -1.07 & -0.69\\
        Median stellar age (Gyr) & 11.8 & 12.6 & 12.1\\
        \hline
        \end{tabular}
\end{table}

\begin{figure*}
    \includegraphics[width=\textwidth]{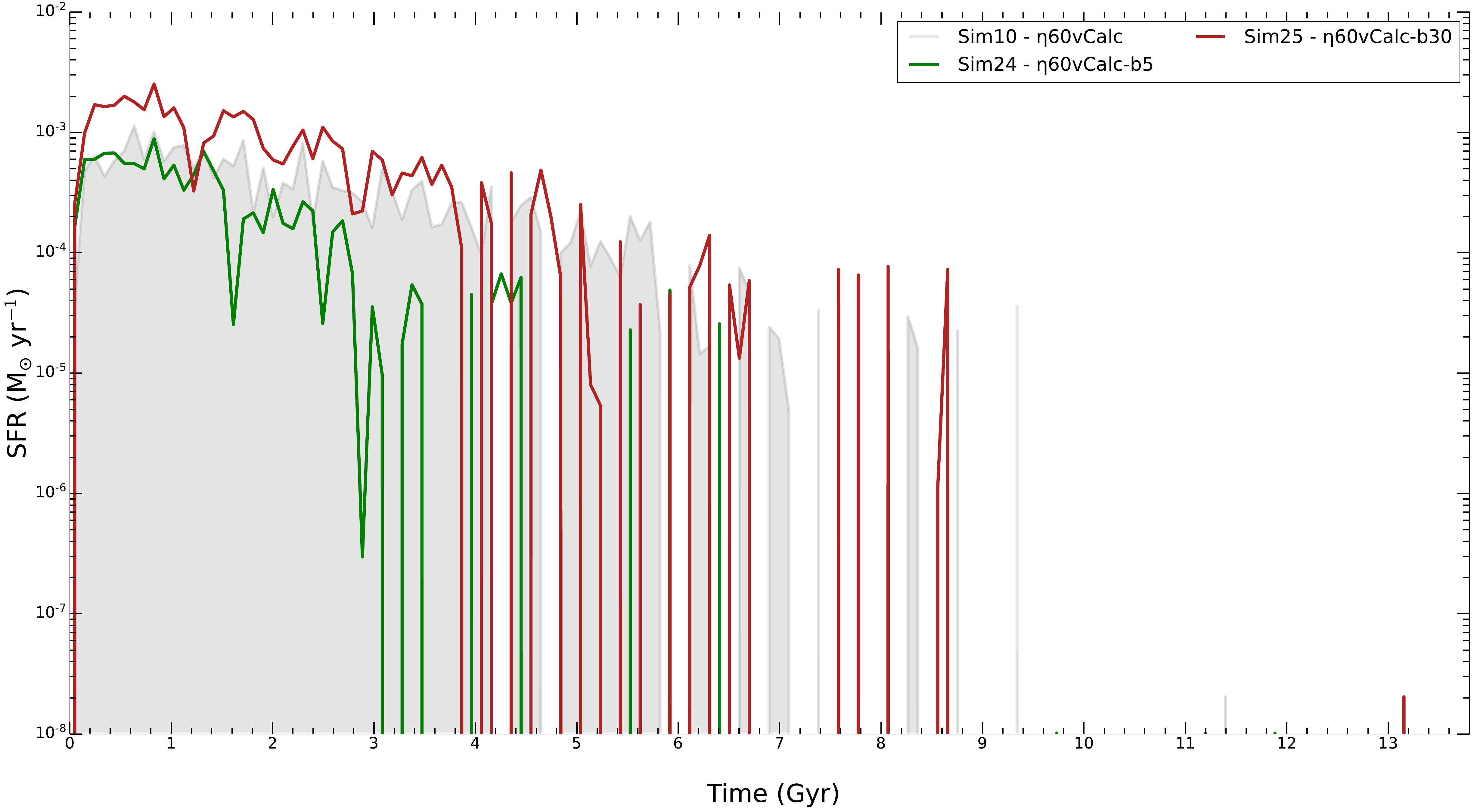}
    \vspace*{2mm}
    \caption{Star formation histories for simulations in under- and over-density environments, compared with the fiducial simulation ($\eta60$vCalc) for Leo II.}
    \label{fig:sfr-baryon}
\end{figure*}

The results in Table~\ref{tab:gas_res} indicates that the simulated galaxy in a lower density environment ($\eta60$vCal-b5) was unable to completely deplete the tidal region in current times, yielding a residual gas mass over 1 dex higher than the predicted upper limit of $2 \times 10^4$ M$_{\odot}$ for Leo II \citep{Grcevich2009}. This outcome can be attributed to the lower star formation rates over time, as depicted in Fig.~\ref{fig:sfr-baryon}. While the stellar mass buildup was reduced by 45$\%$ in this simulation, it still reached an acceptable final stellar mass for Leo II \citep{Kirby2011metals}. Moreover, the measured metallicity of stars represented by [Fe/H] was in better agreement with the reference value of [Fe/H]$_{\text{obs}} = -1.59$ as also reported by \citet{Kirby2011multi}. In this work, the authors further discussed the possibility that Leo II might have spared gravitational interactions capable of gas stripping, in contrast with other dwarfs like Sculptor and Sextans. In this scenario, Leo II could have spent most of its evolutionary time in a low-density region of the Local Group, making it less likely to encounter and accrete the low-metallicity gas reservoir required to explain their interpretation of Leo II´s metallicity distribution function. 

In the simulation in a low-density environment, the galaxy still had adequate gas supply to suport an adequate level of star formation, but still did not match the stellar metallicity expected by \citet{Kirby2011multi} for Leo II. Further reducing the available gas reservoir could improve this metallicity match, but at the expense of a reduced star formation, which would not be consistent with Leo II. Furthermore, the median metallicity obtained in the low-density environment simulation aligns more closely with another reference value of [Fe/H] = $-1.1$ found in the work by \citet{Dolphin2002}. This suggests that Leo II may have resided in a lower-density region for most of its evolution. This difference in environment could partially explain the observed variations in the metallicity constraints. 

The simulation $\eta30$vCal-b30 regarding the evolution of the isolated galaxy in an overdense region yielded no gas within the tidal region at the end of the simulation. The stellar mass formed increased by $90\%$ compared to the fiducial simulation. This trend can be justified by higher values of SFR in Fig.~\ref{fig:sfr-baryon}. The metallicity in stars improved slightly when compared to the observational references. 

In both simulations testing the changes in the baryonic fraction, older stars were generated when compared to the fiducial simulation (+0.8 Gyr in the underdense and +0.3 Gyr in the overdense simulation), which are also higher than the observationally estimated values of 8.8 Gyr \citep{Orban2008} or 9.4 Gyr \citep{Dolphin2002}. The persisted disparities in the ages and stellar metallicities may suggest the need for a different gas infall dynamics. This dynamics could have been characterized by more intermittent or delayed inflows during the galactic evolution of Leo II.

\section{Discussion}   \label{discussion}

Many important features of the dwarf galaxy Leo II were satisfactorily reproduced with isolated SPH simulations, particularly in the chosen fiducial simulation for further analysis (simulation $\eta60$vCalc). However, the remaining tension between observational constraints on stellar metallicity and ages in Leo II and the results of the isolated simulations raises important questions about the evolution of this galaxy. 

An initial assumption is that additional factors contributing to gas depletion, such as environmental influences like tidal stripping and ram pressure, may be required to fully account for all the constraints observed in Leo II. However, as mentioned in the introduction, there are studies in the literature suggesting that Leo II may have undergone an isolated evolution separate from a larger host galaxy \citep{Coleman2007, Koch2007, Lanfranchi&Matteucci2010, Munoz2018}. Moreover, if we consider the data from \citet{Battaglia2022} about a potential last pericentric passage around 2 Gyr ago, the findings for the star formation history in Leo II documented in literature \citep{Dolphin2002, Koch2007, komiyama2007, Lanfranchi&Matteucci2010, Kirby2011multi} and the results of our work show that it would not significantly affect its star formation, which was already quenched before infall (see Fig.~\ref{fig:sfr}).

The simulations with modified initial baryon mass fraction revealed that the initial gas reservoir can have a significant influence on gas depletion, star formation history and stellar metallicity evolution of a simulated dwarf galaxy. This may potentially provide valuable insights into the environmental history of Leo II. Specifically, reducing the initial mean gas density (and thus, the average dynamical timescale) improved the reproduction of stellar metallicity, although a discrepancy of $-33\%$ persisted when compared to the reference value of \citet{Kirby2011multi}.

In a scenario of minimized role of environmental effects, the modelling of the gas infall regime may be the most critical factor influencing the reproduction of stellar metallicity and ages for the Leo II galaxy. It is possible that isolated simulations, parameterized with spherically symmetric gas infall starting at $t = 0$, may not adequately account for the necessary complexity and intermittency of galactic inflows over time, as previously discussed in sections \ref{optsim} and \ref{metallicity}. Furthermore, these discrepancies, coupled with the presence of a super-solar metallicity tail in Fig.~\ref{fig:iron_prof}, may suggest that the dilution of metals in the interstellar medium remains inefficient during the evolution of the simulated galaxy. 

An alternative to alleviate the observed tensions could involve intermittent infalls of more pristine gas over time, as also advocated by \cite{Koch2007} and \cite{Kirby2011multi} for Leo II. An isolated simulation could only reproduce it with ad hoc procedures, which would introduce a level of complexity not suitable for an isolated analysis, since cosmological simulations would offer a self-consistent means to address this issue. Nevertheless, we intend to investigate these effects in an upcoming study that incorporates cosmological simulations. This future work will explore the role and interplay of environmental effects and feedback mechanisms in relatively isolated dwarf spheroidal galaxies similar to Leo II, for comparison with our findings within an isolated framework. 

The source of gas for an intermittent gas infall regime could be supplied by interactions of Leo II with smaller systems over time. For instance, \citet{komiyama2007} observed an extended stellar halo for Leo II, in which they found evidence of a knotty substructure that could be associated with remnants of a globular cluster located around the eastern edge of Leo II. This hypothetical globular cluster may have undergone tidal disruption due to interactions between Leo II and other dwarf galaxies over time (see also discussion in \citet{Lepine2011}). Such interactions could also serve as a source of more pristine gas. As such interactions would not have been continuous in the past, but rather episodic in nature, they could support the hypothesis of an intermittent gas regime, which has been suggested in previous observational studies \citep{Koch2007, Kirby2011multi} and further supported here with SPH simulations. 


Another argument in this regard comes from high-resolution cosmological simulations conducted by \citet{Wheeler2015}, which predicted that subhaloes of isolated dwarf galaxies would give rise to dwarf galaxies as satellites of larger dwarfs. They discovered that certain subhaloes could potentially host ultra-faint dwarfs with stellar masses $M_* \lesssim 10^4$ M$_{\odot}$. Additionally, they estimated, using dark-matter only simulations from the ELVIS suite, that each isolated dwarf galaxy with $M_* \sim 10^6$ M$_{\odot}$ in the Local Group would have a 35\% probability of hosting at least one satellite with $M_* > 3000$ M$_{\odot}$.

An alternative approach to investigate the metallicity evolution in more detail involves testing models that resolve the interstellar medium (ISM) in dwarf galaxies, which have seen development in recent years. In the LYRA project, for instance, \citet{Gutcke2021} employed a high-resolution galaxy formation model using the code AREPO for an idealized isolated dwarf galaxy. Their model resolves the ISM down to 10 K, facilitates the formation of individual stars sampled from the IMF, and accounts for single supernova blast waves with variable energy, achieving a gas mass resolution of 4 M$_{\odot}$. In their conclusions, the authors argue that the metallicity distribution was significantly influenced by the energy injection scheme.

\section{Limitations of this study} 

Due to the inherent nature of isolated galaxy simulations, this work does not delve into the detailed effects of reionization on the gas dynamics of dwarf galaxies \citep[e.g.][]{Wheeler2015, Benitez2020, Rey2020}. Conversely, starting the simulations at a later time (to avoid such effects) would force the inclusion of a pre-existing stellar population, introducing more uncertainties into the input parameters \citep[e.g.][]{Pasetto2010}. 

Another point to consider is that the isolated simulations conducted in this study do not account for the time-dependent mass variations of dark matter halos, which could potentially amplify the impact of winds resulting from kinetic stellar feedback, especially at higher redshifts, due to the reduced gravitational potential \citep[e.g.][]{Sawala2010}. Furthermore, this work did not analyze the influence of a cored dark matter halo for Leo II. However, it is a plausible condition for dwarf spheroidal galaxies, where the star formation processes can act to flatten the central density cusps \citep[e.g.][]{Koch2007, komiyama2007, Pasetto2010}. 

The findings of our study are not immune to limitations imposed by resolution constraints. Initially, our research did not encompass resolving the ISM for our target galaxy, as most of these systems in the Local Group are currently gas-depleted. Our primary aim was to explore the parameter space of the kinetic feedback model for injecting energy into the ISM due to stellar evolution. Nevertheless, addressing this matter in future investigations is crucial for comparing with our intermediate-resolution outcomes. Improving resolution to resolve the smallest molecular clouds and refine the stellar feedback granularity in energy injection could yield more precise insights into the galactic evolution of dwarf spheroidals. It is noteworthy, however, that uncertainties persist regarding the physics at these smaller resolved scales, including the relative importance of magnetic fields, cosmic rays, interstellar radiation, non-equilibrium chemistry, among other factors \citep[e.g.][]{Gutcke2021}.

Finally, this work does not incorporate any specific modifications to the Schmidt law \citep{Kennicutt1998star, Kennicutt1998global} to account for the low-mass regime of dwarf spheroidal galaxies, although there have been efforts in the literature to refine this law to better suit irregular dwarf galaxies \citep[e.g.][]{Roychowdhury2017}.


\section{Summary and Conclusions}

The evolution of an isolated dwarf spheroidal galaxy was investigated using SPH numerical simulations, with Leo II (PGC 34176) of the Local Group as our default model. The Galactocentric distance of Leo II ($235.6^{+13.9}_{-9.14}$ kpc estimated by \citet{Li2021}), places this galaxy as one of the most distant satellites of the Milky Way, making it suitable for studying internal feedback mechanisms with limited influence from environmental effects. 
Our primary objective was to assess whether the inclusion of winds driven by stellar evolution and SNe in numerical simulations would enhance our ability to replicate key observational features and the star formation history of a dwarf spheroidal galaxy. These winds have the potential to influence both inflow and outflow processes in the galaxy, also affecting star formation quenching. Some studies show that the degree of SN feedback induced star-formation suppression is a function of the halo mass and is particularly relevant in the low-mass regime \citep[eg.][]{Springel2003, Pillepich2018, Higgs2021}. 

We employed a modified version of the SPH code GADGET-3, including the following subresolution physics: radiative cooling and heating from photoionizing background, multiphase model for star formation, stellar evolution processes, chemical enrichment for 11 elements and stellar feedback in both thermal and kinetic forms \citep{Springel2003, Tornatore2007}. The primary free parameters and inputs in our model included the mass loading factor and velocity of SN-driven winds, fraction of the supernovae energy carried by the winds, binary fraction of stars, stellar initial mass function, and initial baryonic mass fraction of the galaxy. 

The optimal simulation to model Leo II was obtained for the parameters $\eta = 60$ and v$_{\text{wind}} = f(\eta$). Several key properties of this simulation, including the total mass of stars formed, the approximate duration of the star formation, the value of the mass loading factor, the residual gas mass within the tidal radius, and the total mass within 600 pc agreed with related observational constraints found in the literature. 

However, there were challenges in replicating the median stellar metallicity and stellar ages in the simulations, which raised questions about the interplay of cosmic gas infall with stellar evolution which happened during the lifetime of Leo II. A primary hypothesis considered was that the simple isotropic gas infall could not adequately capture the intermittent and more pristine gas infall onto the galaxy, possibly necessary to reproduce the observed metallicity patterns identified in spectroscopic analyses of red giants and chemical modelling of Leo II \citep{Dolphin2002, Koch2007,Lanfranchi&Matteucci2010, Kirby2011multi, Kirby2011metals}. Exploring variations in the IMF, binary fraction and DM halo mass did not lead to improvements in reproducing these constraints.  

Modifying the initial gas reservoir mass and density by adjusting the baryonic mass fraction resulted in variations in the simulation outcomes. Specifically, the simulation with a lower-density environment produced median stellar metallicities in better agreement with the observational constraints, although there were still discrepancies in terms of gas depletion and stellar ages. Indirectly, modifying such a parameter is equivalent to changing the average dynamical timescale, since $t_{\text{dyn}} \sim1 / \sqrt{G\overline{\rho}}$. In these simulations, the gas falls more slowly to the center of the galaxy, leading to a more effective dilution of metals in the interstellar medium, since the reduction in the stellar mass alone could not account for the reduction in stellar metallicity in the simulations. However, despite these improvements, the stellar ages were not adequately reproduced. Therefore, these results suggest that achieving an even slower gas infall, combined with intermittency, might be crucial in accurately reproducing the metallicity patterns observed in Leo II. 

From a broader perspective, the results obtained thus far suggest that currently quenched dwarf galaxies may not necessarily need to evolve within clusters or groups of galaxies to exhibit some of the typical characteristics observed in local dwarf spheroidal galaxies. This inference suggests that stellar feedback alone might be a sufficient factor in shaping at least some of these systems as we observe them today, when environmental effects are minimized. 

However, it is important not to overlook the complexity of the gas infall process needed to accurately reproduce critical features such as stellar metallicities and ages, even for galaxies that do not show evident signs of tidal effects or disruptions over time, as is the case of the dSph Leo II. The findings of this work suggest that it is not only the outflows resulting from supernovae that are essential for explaining their SFH and evolutionary trajectory, but also that the gas inflow regime may have played a pivotal role in shaping such systems. In the case of Leo II, matching both its observed stellar metallicity while also maintaining a sufficient level of star formation activity has proven to be a complex challenge in isolated simulations. 

In the observational context, a larger sample of isolated dwarf galaxies from surveys conducted by future facilities such as the Vera Rubin observatory and Extremely Large Telescope (ELT) holds the potential to provide more definitive insights into the question of whether dwarf galaxies quenched solely by supernovae exist in the universe, as discussed by \citet {Collins&Read2022}.

Attempting to forcibly replicate all the expected observational constraints for the galaxy Leo II by introducing ad hoc mechanisms to mimic intermittent gas inflows in isolated SPH simulations could result in overly artificial modeling, particularly considering that we are working with sub-resolution models for star formation and feedback. This was not the primary intention of this study. Instead, the main objective was to assess whether a combination of thermal and kinetic stellar feedback, with a reasonable choice of parameters, could produce simulated galaxies resembling quenched dwarf spheroidal galaxies that evolved in more isolated environments. In this regard, this work has provided further numerical simulation-based evidence that stellar feedback might be sufficient to reproduce many aspects of such galaxies. However, the intricacies of the cosmic gas inflows and outflows could be more precisely addressed in cosmological simulations (one of the future steps of our study to be presented elsewhere), which could provide a more comprehensive picture of these complex processes. 


\section{Acknowledgments}
We would like to acknowledge the Coordenação de Aperfeiçoamento de Pessoal de Nível Superior - Brasil (CAPES) - Finance Code 001 for the financial support provided for this research. Additionally, we acknowledge the National Laboratory for Scientific Computing (LNCC/MCTI, Brazil) for granting access to HPC resources on the Santos Dumont supercomputer. We thank Volker Springel for providing us with a preliminary version of the GADGET-3 code and the code to generate initial conditions. Finally, we thank the anonymous reviewers for their insightful comments and constructive feedback, which significantly contributed to enhancing the quality and clarity of this manuscript. 

%




\pagebreak
\section*{Data Availability}
 
The simulation data is available upon request.

\appendix

\section{Numerical convergence}

To investigate how the adopted resolution could affect our results, we compared the galactic evolution using simulations with varying numbers of SPH particles. These tests involved simulations at lower resolution (with half the number of particles) and higher resolution (with twice the number of particles). The initial setup and all parameters, except for $\eta$ and $v_{\text{wind}}$, remained the same in all the simulations. Table~\ref{tab:resolution} summarizes selected results for the different runs, where $m_{\text{dm}}$ and $m_{\text{gas}}$ are the dark matter particle mass and initial gas particle mass, respectively. Tests were also performed with ten times more SPH particles, but it led to stellar particle masses lower than $10^3$ M$_{\odot}$, which could compromise the IMF representation at such high mass resolution. In this regime, stochastic variations in stellar populations can become significant and affect the stellar feedback \citep[e.g.][]{Smith2021}. 

\begin{table*}[h]
	\centering
        \caption{Influence of mass resolution in the simulations.}
        \label{tab:resolution}
        \resizebox{\textwidth}{!}{\begin{tabular}{ccccccccc} 
		\hline
	    Simulation & Gas particles & $m_{\text{dm}}$ (M$_{\odot}$) & $m_{\text{gas}}$ (M$_{\odot}$) & $\eta$ & $v_{\textit{}{\text{wind}}}$ (km/s) & Residual tidal gas mass (M$_{\odot}$) & Stellar mass (M$_{\odot}$) & $[\text{Fe/H}]_{\text{star}}$\\
		\hline
		$\eta60$vCalc & 20000 & 5.3 $\times 10^4$ & 1.6 $\times 10^4$ & 60 & 96 & 0 & 2.0 $\times 10^6$ & -0.51\\
        lower resolution & 10000 & 1.1 $\times 10^5$ & 3.2 $\times 10^4$ & 60 & 96 & 7.6 $\times 10^5$ & 9.9 $\times 10^5$ & -0.85\\
        lower resolution & 10000 & 1.1 $\times 10^5$ & 3.2 $\times 10^4$ & 70 & 89 & 1.3 $\times 10^6$ & 9.5 $\times 10^5$ & -0.86\\
        lower resolution & 10000 & 1.1 $\times 10^5$ & 3.2 $\times 10^4$ & 75 & 86 & 7.8 $\times 10^5$ & 1.4 $\times 10^6$ & -0.64\\ 
        lower resolution & 10000 & 1.1 $\times 10^5$ & 3.2 $\times 10^4$ & 80 & 65 & 3.0 $\times 10^6$ & 1.0 $\times 10^7$ & -0.04\\
        higher resolution & 40000 & 2.7 $\times 10^4$ & 8.0 $\times 10^3$ & 60 & 96 & 3.1 $\times 10^6$ & 1.0 $\times 10^7$ & 0\\
        higher resolution & 40000 & 2.7 $\times 10^4$ & 8.0 $\times 10^3$ & 50 & 105 & 0 & 3.4 $\times 10^6$ & -0.29\\
        higher resolution & 40000 & 2.7 $\times 10^4$ & 8.0 $\times 10^3$ & 45 & 111 & 0 & 3.4 $\times 10^6$ & -0.28\\
        higher resolution & 40000 & 2.7 $\times 10^4$ & 8.0 $\times 10^3$ & 35 & 125 & 0 & 1.4 $\times 10^6$ & -0.76\\
        \hline
        \end{tabular}}
\end{table*}

\begin{figure*}[h]
    \includegraphics[width=\textwidth]{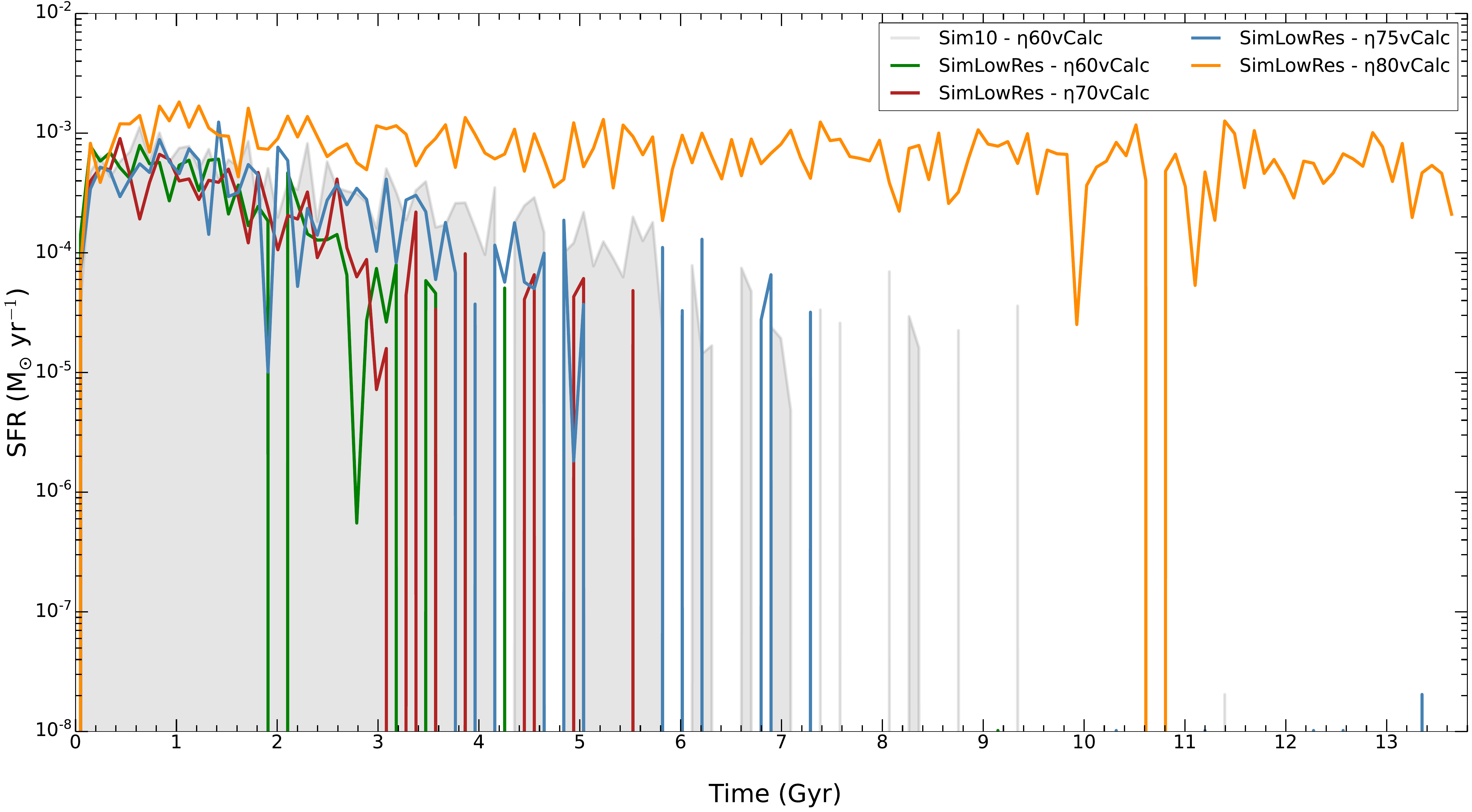}
    \vspace*{-5mm}
    \caption{SFH of simulations at lower resolution.}
    \label{fig:sfr-lowres}
\end{figure*}

\begin{figure*}[h]
    \includegraphics[width=\textwidth]{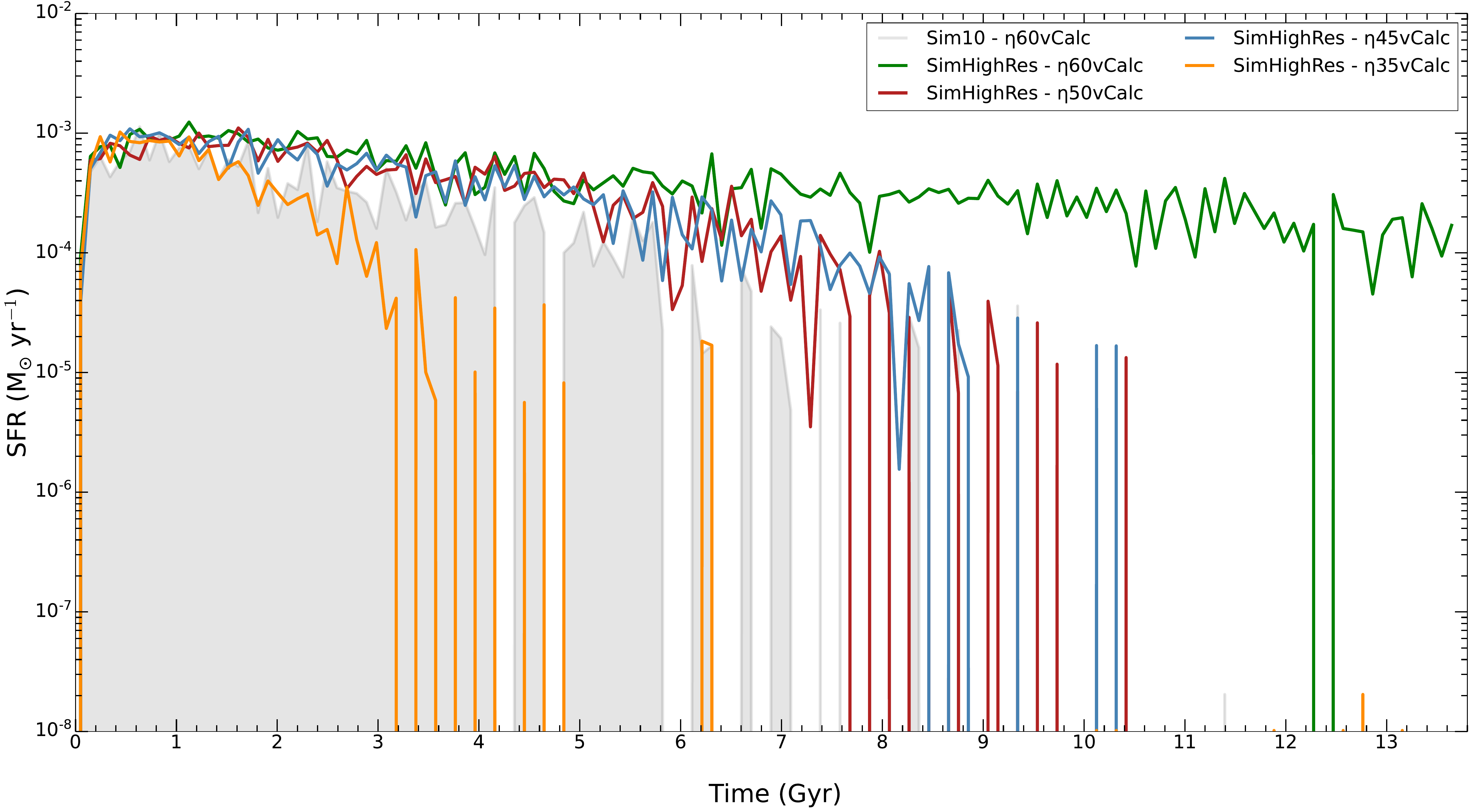}
    \vspace*{-5mm}
    \caption{SFH of simulations at higher resolution.}
    \label{fig:sfr-highres}
\end{figure*}

Figs.~\ref{fig:sfr-lowres} and \ref{fig:sfr-highres} display the star formation histories of the SPH simulations conducted at lower and higher resolutions, in comparison to the fiducial simulation $\eta60$vCalc (gray-shaded in these figures). The equivalent models, with the same choice of the loading factor $\eta$, are the ones depicted in green. While we do not achieve numerical convergence in a strict sense, consistent results for the SFH were obtained across both higher and lower resolutions, with variations in $\eta$ by factors lower than approximately 1.5.

In the lower-resolution tests, Table~\ref{tab:resolution} indicates that complete gas depletion within the tidal radius was not achieved, with gas masses ranging between $10^5$ and $10^6$ M$_{\odot}$ in this region. The simulation most similar to the fiducial case was the one implemented with $\eta = 75$ (see also Fig.~\ref{fig:sfr-lowres}). When the fiducial wind mass loading factor $\eta = 60$ was used, there was a $50 \%$ reduction in the final stellar mass. To obtain similar results at lower resolution, it was necessary to increase the value of $\eta$. The optimal $\eta$ value appeared to fall within the range of $\eta \sim75-80$ to reproduce a similar SFH (Fig.~\ref{fig:sfr-lowres}), final stellar mass and metallicity. Different choices of $\eta$ did not result in significant variations in the star formation peak, which remained around $10^{-3}$ M$_{\odot}$ yr$^{-1}$. The stellar metallicity values obtained were similar to those observed at the regular resolution used throughout this work. 

In the higher-resolution tests, Table~\ref{tab:resolution} indicates that the gas depletion within the tidal radius was achieved for $\eta = 35-50$. The simulation most similar to the fiducial case was the one implemented with $\eta = 35$ (see also Fig.~\ref{fig:sfr-highres}). When the fiducial wind mass loading factor $\eta = 60$ was used, an increment of approximately five times was observed in the final stellar mass. Additionally, a residual gas mass of approximately $10^6$ M$_{\odot}$ was observed within the tidal radius. Consequently, it was necessary to decrease the value of $\eta$ to achieve similar results at higher resolution.  The optimal $\eta$ value appeared to be within the range of $\eta \sim35-45$ to reproduce a similar SFH (Fig.~\ref{fig:sfr-highres}), final stellar mass and metallicity. The stellar metallicity values obtained were similar to those observed at the regular resolution.  

In combination, the results of the numerical convergence tests indicate that the selection of optimal stellar feedback parameters for simulating the galaxy Leo II is dependent on the resolution. Furthermore, the convergence with respect to gas depletion was achieved between the regular and higher resolutions.

\pagebreak


\bibliography{references}{}
\bibliographystyle{aasjournal}



\end{document}